\documentclass[reprint,amsmath,amssymb,aps]{revtex4-1}

\usepackage{graphicx}
\usepackage{dcolumn}
\usepackage{bm}
\usepackage{bbold}
\usepackage{float}

\usepackage{latexsym}
\usepackage{amssymb}
\usepackage{mathrsfs}
\usepackage{amsfonts}
\usepackage{slashed}

\usepackage[utf8]{inputenc}
\usepackage{graphicx}
\usepackage{epsfig}
\usepackage{amsmath}
\usepackage{color}

\allowdisplaybreaks
\newcommand{\lsim}{\stackrel{<}{_\sim}}
\newcommand{\gsim}{\stackrel{>}{_\sim}}

\begin{document}
\title{Lepton Flavour Violation in Hadron Decays of the Tau Lepton  \\ in the Simplest Little Higgs Model}

\author{A. Lami}
\email{Andrea.Lami@ific.uv.es}
\author{J. Portol\'es}
\email{Jorge.Portoles@ific.uv.es}
\affiliation{Instituto de F\'{\i}sica Corpuscular, Universitat de Val\`encia - CSIC, Apt. Correus 22085, E-46071 Val\`encia, Spain}
\author{P. Roig}
\email{proig@fis.cinvestav.mx}
\affiliation{Departamento de F\'{\i}sica, Centro de Investigaci\'on y de Estudios Avanzados del Instituto Polit\'ecnico
Nacional, Apartado Postal 14-740, 07000 M\'exico D.F., M\'exico}

\date{\today}

\begin{abstract}
We study Lepton Flavour Violating hadron decays of the tau lepton within the Simplest Little Higgs model. Namely we consider 
$\tau \rightarrow \mu (P, V, PP)$ where $P$ and $V$ are short for a pseudoscalar and a vector meson. We find that, in the
most positive scenarios, branching ratios for these processes are predicted to be, at least, four orders of magnitude smaller than present 
experimental bounds. 
\begin{description}
\item[PACS numbers]: 11.30.Hv, 12.60.Cn, 13.35.Dx
\item[Keywords]: Lepton Flavour Violation, Hadron Tau Decays
\end{description}
\end{abstract}

\maketitle
\section{Introduction}
The experimental observation \cite{Agashe:2014kda} that neutrinos are massive and oscillate between flavours indicates that Lepton Flavour Violation (LFV) does
take place in the neutral sector. When neutrino mass terms are included in the Standard Model (SM) they induce also one-loop LFV decays of charged leptons (CLFV) like, for instance, $\mu \rightarrow e \gamma$. However due to the tiny ratio between the neutrino mass and the electroweak energy scale, 
$B(\mu \rightarrow e \gamma) \lsim 10^{-54}$ \cite{Marciano:1977wx,Bilenky:1977du,Cheng:1977nv}. Its present upper bound (at $90 \, \% \, C.L.$) is given by the
MEG experiment $B(\mu^+ \rightarrow e^+ \gamma) \leq 5.7 \times 10^{-13}$ \cite{Adam:2013mnn}, expecting to reach one order of magnitude less in the current
upgrade. Hence the SM predicts unobservable branching ratios for CLFV decays in the foreseen future.  
\par
This setting provides an ideal benchmark for the hunt of New Physics beyond SM. The experimental observation of CLFV is the goal of a bunch of excellent 
dedicated experiments \cite{CeiA:2014wea} like, for instance, MEG, MEGA, SINDRUM and Mu3e in the search of muon decays, and those looking for muon conversion in the
presence of nuclei, SINDRUM II, Mu2E or COMET/PRISM. The first generation of B-factories, that stand for $\tau$ factories too like BaBar or Belle, have joined in the
pursuit of CLFV decays coming from the $\tau$ lepton \cite{CeiA:2014wea}. Though they have focused mainly in processes with leptons (and photons) in the
final state, both experiments have also provided excellent bounds on hadron decays of the tau lepton, for the first time \cite{Hayasaka:2009zza,Vasseur:2015vga,Miyazaki:2012mx}, for instance $\tau \rightarrow \mu P$, $\tau \rightarrow \mu V$, $\tau \rightarrow \mu P P$, where
$P (V)$ is short for a pseudoscalar (vector) meson. The study of LFV in decays of the tau lepton are also one of the main goals of the future
SuperKEKB/Belle II project under construction at KEK (Japan) \cite{Shwartz:2015kja}. 
\par
While the theoretical study of LFV tau decays involving only leptons has a long story (see \cite{Vicente:2015cka} and references therein), LFV hadron decays of the tau lepton have only been slightly surveyed \cite{Pich:2013lsa}. There are a few studies within models of SUSY 
\cite{Brignole:2004ah,Fukuyama:2005bh,Arganda:2008jj,Herrero:2009tm}, within the Littlest Higgs model with T-parity \cite{Liu:2009su,Goto:2010sn}, and
with the 331 model \cite{Hua:2014yna}. A thorough study of the role of the Higgs contribution to the decay of $\tau \rightarrow \mu \pi \pi$ has also
been carried out \cite{Celis:2013xja}.  Here we explore those decays within the Simplest Little Higgs (SLH) model \cite{Kaplan:2003uc,Schmaltz:2004de}.
\par
Little Higgs (LH) models \cite{ArkaniHamed:2002qy,ArkaniHamed:2002qx,Schmaltz:2005ky,Perelstein:2005ka} arise from the old idea of a composite Higgs boson \cite{Georgi:1984af,Dugan:1984hq} where some collective symmetry breaking, that allows the Higgs mass to become loop suppressed, has been implemented \cite{Bellazzini:2014yua}. As a consequence electroweak symmetry breaking is fulfilled by a naturally light Higgs sector, and the discovery of the Higgs boson with a relatively light mass 
$M_h \simeq 125 \, \mbox{GeV}$  \cite{Aad:2012tfa,Chatrchyan:2012xdj} could substantiate a little Higgs model. General features of a composite Higgs 
involve: i) a scale of compositeness $f$; ii) a hierarchy between the electroweak (Higgs vev $v$) and the compositeness scale, i.e. $v/f \ll 1$; iii) a 
Higgs potential that is (entirely or in part) radiatively generated. Different composite Higgs models differ, essentially, on which and how many pieces
in the potential are radiatively generated \cite{Bellazzini:2014yua}. Little Higgs models, in particular, are characterized generically by a loop-level generated mass (that accords with its smallness) and a tree-level generated quartic coupling. Hence some tuning has to be introduced in order to balance both scales. 
In addition they contain new \lq \lq little" particles, with masses around the scale of compositeness $f \sim
1 \, \mbox{TeV}$, that cancel one-loop quadratically divergent contributions to the Higgs mass from Standard Model loops. One expects the need of a more fundamental theory, an ultraviolet completion, at a scale of $\Lambda_f \sim 4 \pi f \sim 12 \, \mbox{TeV}$ where the description given by the LH models may become strongly coupled.  
\par 
Although the Little Higgs mechanism can be implemented in different ways giving diverse models, these can be grouped into two types that share
many common phenomenological features. 
LH models can be categorized into two classes depending on the way the Standard Model $SU(2)_L$ group is inserted
\cite{Kaplan:2003uc,Han:2005ru}: {\em product group} models where the electroweak group arises from the diagonal breaking of a product gauge group, as the Littlest Higgs \citep{ArkaniHamed:2002qy}, and {\em simple} models  when the SM $SU(2)_L$ embedding happens through the breaking of a simple group, as is the case of the
SLH \cite{Kaplan:2003uc,Schmaltz:2004de}. A common feature to all LH models is their extended spectrum of gauge bosons and fermions, playing the latter a 
crucial role in the implementation of the collective symmetry breaking and, hence, in the cancellation of non-wished ultraviolet divergences. Moreover
they become an important asset as possible signals to discern between different models.
\par
LHC Higgs data have already challenged predictions of the LH models \cite{Han:2013ic,Kalyniak:2013eva} and it was soon pointed out the existence of a possible stress with the first measurements of diphoton decays of the Higgs boson. However later measurements of this process \cite{Aad:2014eha,Khachatryan:2014ira} have eased the tension.
\par
In the next section we collect, for completeness, the characteristics, properties and features, of the SLH model. In Section III we proceed to the explain the
calculation of the LFV hadron tau decays $\tau \rightarrow \mu P$, $\tau \rightarrow \mu V$, $\tau \rightarrow \mu P P$ in the SLH model. We will also detail
the procedure of hadronization. The results and their discussion will be postponed to Section IV. Finally the conclusions of our work are given in Section V.

\section{The Simplest Little Higgs Model}
The SLH model \cite{Kaplan:2003uc,Schmaltz:2004de,Han:2005ru} is constructed by embedding the electroweak SM gauge group $SU(2)_L \otimes U(1)_Y$ into a  
$SU(3) \otimes U(1)_X$ gauge group. The collective symmetry breaking procedure is realized through two complex scalar fields $\Phi_{1,2}$, which are triplets under $SU(3)$:
\begin{equation} \label{eq:kh12}
{\cal L}_{\Phi} = \left( D_{\mu} \Phi_1 \right)^{\dagger} D^{\mu} \Phi_1 + \left( D_{\mu} \Phi_2 \right)^{\dagger} D^{\mu} \Phi_2 \, .
\end{equation}
Their initial scalar potential has a $[SU(3) \otimes U(1)]^2$ global symmetry that breaks spontaneously to $[SU(2) \otimes U(1)]^2$, with corresponding
vacuum expectation values given by $f_{1,2} \sim {\cal O }(1 \, \mbox{TeV})$ and yielding five Nambu-Goldstone bosons from each scalar. Meanwhile the diagonal subgroup of the $[SU(3) \otimes U(1)]^2$, i.e. $SU(3)_L \otimes U(1)_X$, that has been gauged, breaks down to $SU(2)_L \otimes U(1)_Y$ via the 
$\langle \Phi_{1,2} \rangle$ vacuum condensates. 
Here the hypercharge group $U(1)_Y$ is identified with the unbroken linear combination of the $U(1)$ and the eighth generator of $SU(3)$. Notice that this model has no custodial symmetry \cite{chang:2003un,chang:2003zn} (see below in this Section).
\par
The scalar multiplets are given by a non-linear sigma model. They include the SM Higgs as well as new Goldstone bosons:
\begin{eqnarray} \label{eq:phi12}
\Phi_1 & = & \exp \left( i \frac{\Theta'}{f} \right) \, \exp \left( i t_{\beta} \frac{\Theta}{f} \right)  \, \left( \begin{array}{c}
                                                                                                                      0 \\
                                                                                                                      0 \\
                                                                                                                      f \, c_{\beta}
                                                                                                                    \end{array} \right) , \nonumber \\
\Phi_2 & = & \exp \left( i \frac{\Theta'}{f} \right) \, \exp \left( - \frac{i}{t_{\beta}} \frac{\Theta}{f} \right)  \, \left( \begin{array}{c}
                                                                                                                      0 \\
                                                                                                                      0 \\
                                                                                                                      f \, s_{\beta}
                                                                                                                    \end{array} \right) . 
\end{eqnarray}
Here $t_{\beta} \equiv \tan \beta = f_1/ f_2$, $s_{\beta} \equiv \sin \beta$, $c_{\beta} = \cos \beta$  and $f^2 = f_1^2 + f_2^2$. In Eq.~(\ref{eq:phi12}),
$\Theta'$ and $\Theta$ carry the Goldstone bosons. The term $\exp \left( i \Theta' /f \right)$ can be rotated away through a $SU(3) \otimes U(1)_X$ gauge
transformation (unitary gauge) and 
\begin{equation} \label{eq:theta}
\Theta = \left( \begin{array}{cc}
                \mathbb{0}_{\scriptscriptstyle 2 \times 2}  & h \\
                h^{\dagger} &  0 
                \end{array} \right) \, + \, \frac{\eta}{\sqrt{2}} \, \mathbb{1}_{\scriptscriptstyle 3 \times 3} \, ,
\end{equation}
that includes the complex Higgs doublet $h \equiv \left( h^0, h^- \right)^T$ and the scalar singlet $\eta$. Upon electroweak symmetry breaking we will have
\begin{equation} \label{eq:ewsb}
 h = \exp \left( \frac{i}{v} \, \chi^j \tau_j \right) \left( \begin{array}{c}
                                                                      \frac{1}{\sqrt{2}} \left( v + H \right) \\
                                                                      0
                                                                      \end{array} \right).
\end{equation}
being $H$ the SM Higgs and $\langle h^0 \rangle = v / \sqrt{2} \simeq 0.174 \, \mbox{TeV}$. Hence, in the unitary gauge, the three $\chi_j$ degrees of freedom can be gauged away. They provide
the longitudinal components of the SM gauge bosons. The scalar singlet $\eta$ plays no role in the following (see however Ref.~\cite{Han:2005ru}).
\par 
Let us now consider the fermion sector. The SM doublets of leptons and quarks have to be expanded into $SU(3)$ left-handed triplets where new fermions also appear
and the corresponding $SU(3)$ singlet right-handed fermions are also added:
\par
- \underline{Leptons}. There is a new heavy neutrino $N_k$ but there is no right-handed light neutrino. As a consequence light neutrinos have no mass:
\begin{equation} \label{eq:leptoN}
L_k^{-1/3} = \left( \nu_k, \ell_k, i \, N_k \right)_L^T , \qquad \ell_{k  R}^{-1} ,   \qquad N_{k  R}^0 ,
\end{equation}
with $k=1,2,3$ the family number, and the superscript indicates the $U(1)_X$ hypercharge $y_x$.
\par
- \underline{Quarks}. Contrarily to $SU(2)_L$, $SU(3)_L$ triplets are not free from the triangle anomaly. This does not affect the SM and a possible ultraviolet completion could fix the problem; however we prefer to keep the SLH model free of anomalies. A solution arises by treating asymmetrically the first two families from the third one  \cite{Kong:2004cv}: while the latter is put into the {\bf 3} $SU(3)$ representation, the first two generations of quarks are put into $\overline{\mbox{\bf 3}}$ representations. This is called the {\em anomaly-free} embedding scheme for the three families:
\begin{eqnarray} \label{eq:quarkS}
Q_1^0 \; \; & = & \left( d, -u, i \, D \right)_L^T,\qquad \; d_R^{-1/3}, \; \; u_R^{2/3}, \; \;  D_R^{-1/3}, \nonumber \\
Q_2^0 \; \; & = & \left( s, -c, i \, S \right)_L^T, \qquad \; \;  s_R^{-1/3}, \; \;  c_R^{2/3}, \; \; S_R^{-1/3}, \nonumber \\
Q_3^{1/3} & = & \left( t, b, i \, T \right)_L^T, \qquad \;\; \; \; \;  b_R^{-1/3} , \; \;  t_R^{2/3}, \; \;  T_R^{2/3},  
\end{eqnarray}
where, again, the superscripts indicate the value of $y_x$. A heavy fermion, namely $D$, $S$ and $T$, has been added to each family. 
\par 
The covariant derivative in Eq.~(\ref{eq:kh12}) is given by:
\begin{equation} \label{eq:dcov}
D_{\mu} = \partial_{\mu} \, - \, i \, g \, A_{\mu} \, +  \, i \, g_x \, y_x \, B_{\mu}^x \, , 
\end{equation}
where $g_x = g \, t_{\mbox{\tiny{W}}} / \sqrt{1-t_{\mbox{\tiny{W}}}^2/3}$ being $t_{\mbox{\tiny{W}}} \equiv \tan \theta_{\mbox{\tiny{W}}}$ 
and $\theta_{\mbox{\tiny{W}}}$ the SM weak angle. In Eq.~(\ref{eq:dcov}) $g$ is the SM $SU(2)_L$ coupling and $y_x$ the $U(1)_X$ hypercharge ($y_x = -1/3$ for 
both $\Phi_i$ scalar fields). Observe that as the SM sector is embedded naturally into the larger group, the corresponding gauge couplings of the latter are given altogether by the known SM parameters. 
\par 
The $SU(3)$ gauge bosons read:
\begin{equation} \label{eq:gaubo}
A_{\mu} = A_{\mu}^3 \frac{\lambda^3}{2} + A_{\mu}^8 \frac{\lambda^8}{2} + \frac{1}{\sqrt{2}} \left( \begin{array}{ccc}
                                                                                               0 & W^+ & Y^0 \\
                                                                                               W^- & 0 & W'^{-} \\
                                                                                               Y^{0 \dagger} & W'^{+} & 0
                                                                                               \end{array} \right)_{\mu} ,
\end{equation}
where $\lambda^i$ are the Gell-Mann matrices. The new \lq \lq little" gauge bosons are given by a complex $SU(2)_L$ doublet $( Y_{\mu}^0 , W_{\mu}'^{-} )$
and a $Z_{\mu}'$ boson that arises as a linear combination of $A_{\mu}^8$ and $B_{\mu}^x$. The masses of the new gauge bosons arise from the spontaneous
symmetry breaking of the underlying $[SU(3) \otimes U(1)]^2$ global symmetry and are, accordingly, proportional to the high scale $f$.
For instance:
\begin{eqnarray} \label{eq:gmass}
M_{\mbox{\tiny W'}} & \simeq & \frac{g \, f}{\sqrt{2}} \left( 1 - \frac{v^2}{4 f^2} \right) \, , \nonumber \\
M_{\mbox{\tiny Z'}} & =  & g \, f \, \sqrt{\frac{2}{3 - t_{\mbox{\tiny{W}}}^2}} \left( 1 - \frac{3 - t_{\mbox{\tiny{W}}}^2}{c_{\mbox{\tiny{W}}}^2} \, 
\frac{v^2}{16 f^2} \right) .  
\end{eqnarray}
The quadratic couplings of the Higgs with one heavy and one SM gauge boson induce, after the electroweak symmetry breaking, a mixing between them. In the SLH model this only affects to the
definition of the $Z$ and $Z'$ bosons: $Z' \rightarrow Z' + \delta_Z Z$, $Z \rightarrow Z - \delta_Z Z'$ where, at leading order in the $v/f$ expansion:
\begin{equation} \label{eq:dz}
\delta_Z = \frac{1 -t_{\mbox{\tiny{W}}}^2}{8 c_{\mbox{\tiny{W}}}} \, \sqrt{3 - t_{\mbox{\tiny{W}}}^2} \, \frac{v^2}{f^2} \, . 
\end{equation}
\par
The pure gauge and gauge-lepton Lagrangians are given by:
\begin{equation} \label{eq:vpsi}
{\cal L}_V + {\cal L}_{\psi} \, = \, - \,  \frac{1}{2} Tr \left( G_{\mu \nu} G^{\mu \nu} \right) \,  + \,  \overline{\psi}_k \, i \, \slashed{D} \, \psi_k \, , 
\end{equation}
where $G_{\mu \nu} = (i/g) \left[D_{\mu}, D_{\nu} \right]$ and $\psi_k = \left\{ L_k, \ell_{k R}, N_{k  R} \right\}$, the covariant derivative $D_{\mu}$
being given in Eq.~(\ref{eq:dcov}). The gauge-quark sector is more complicated because of the anomaly-free embedding structure. It reads:
\begin{eqnarray} \label{eq:vquark}
\! \! \! \! \! \! \! \! \! \! {\cal L}_q & = & \overline{Q}_k \, i  \slashed{D}_k \, Q_k \, + \, \overline{q_u}_R \, i  \slashed{D}^u \, q_{u  R} \, + \, \overline{q_d}_R \, i  \slashed{D}^d \, q_{d  R}  \nonumber \\
&& + \, \overline{T}_R \, i  \slashed{D}^u \, T_R \, + \,  \overline{D}_R \, i  \slashed{D}^d \, D_R \, + \,  \overline{S}_R \, i  \slashed{D}^d \, S_R \, , 
\end{eqnarray}
with $q_u = \left\{ u,c,t \right\}$, $q_d= \left\{d,s,b \right\}$ and, remembering that the triplets of the two first families are in the anti-fundamental representation:
\begin{eqnarray} \label{eq:dcoq}
D_{\left\{1,2 \right\} \, \mu} & = & \partial_{\mu} \, + \, i \, g \, A_{\mu}^* \, , \nonumber \\
D_{3 \,\mu} & = & \partial_{\mu} \, - \, i \, g \, A_{\mu}\, + \, \frac{i}{3} \, g_x \, B_{\mu}^x \, , \nonumber \\
D_{\mu}^u & = & \partial_{\mu} \, + \, i \, \frac{2}{3} \, g_x \, B_{\mu}^x \, , \nonumber \\
D_{\mu}^d & = & \partial_{\mu} \, -  \, \frac{i}{3} \, g_x \, B_{\mu}^x \, ,
\end{eqnarray}
where $A_{\mu}$ is given in Eq.~(\ref{eq:gaubo}). 
\par

\par 
The Yukawa sector of the SLH model collects the structure of flavour of the theory.  The lepton masses are generated by:
\begin{equation} \label{eq:yukal}
{\cal L}_Y = i \, \lambda_N^k \, \overline{N}_{k R} \Phi_2^{\dagger} \, L_k \, + \, i \, \frac{\lambda_{\ell}^{kl}}{\Lambda} \, 
\overline{\ell}_{k  R} \, \varepsilon_{m\, n\, p}  \, \Phi_1^m \, \Phi_2^n \, L_l^p \, + \, h.c. \, , 
\end{equation}
where $m,n,p$ are $SU(3)$ indices, and $k,l$ are generation indices. Notice that $\lambda_N$ has been taken diagonal. However $\lambda_{\ell}$ does
not need to be aligned. Upon diagonalization of the latter, the redefined fields of the light leptons $\psi_{k L} = V_{\ell}^{kj} \psi_{j L}$,
for $\psi = \left\{\nu , \ell \right\}$, get a definite mass.
By expanding Eq.~(\ref{eq:yukal}) one also observes a mixing term between heavy and light neutrinos. We separate them by rotating the left-handed sector only and,
up to ${\cal O}(v^2/f^2)$, the physical states for the neutrinos are: 
\begin{equation} \label{eq:rotanu}
\left( \begin{array}{c}
              \nu_i \\
              N_i
              \end{array} \right)_L \, \longrightarrow  \, \left( \begin{array}{cc}
                                                                  1- \frac{\delta_{\nu}^2}{2} & - \delta_{\nu} \\
                                                                  \delta_{\nu} & 1- \frac{\delta_{\nu}^2}{2}
                                                                  \end{array} \right)
                                                                  \left( \begin{array}{c}
                                                                  V_{\ell}^{ij} \nu_{j} \\
                                                                  N_i 
                                                                   \end{array} \right)_L \, ,
\end{equation}
where
\begin{equation} \label{eq:deltanu}
\delta_{\nu} \, = \, - \frac{1}{\sqrt{2} \, t_{\beta}} \, \frac{v}{f} \, . 
\end{equation}
The heavy neutrino masses are given by $m_{N_i}   =  f  s_{\beta} \lambda_N^i$. 
\par
Yukawa quark couplings are rather involved due to the mixing between the heavy \lq \lq little" quarks and the SM ones (we refer the reader 
to Ref.~\citep{Han:2005ru} for a detailed account). As in this article we are only interested in lepton flavour violating processes we will assume,
in the following, no flavour mixing in the quark sector. Accordingly the heavy-light mixing will stay within each family. The proper redefinition of the physical (massive) and left-handed fields is given by: $P_L \rightarrow P_L + \delta_p p_L$ 
and $p_L \rightarrow p_L - \delta_p P_L$ for $P = \left\{T,D,S \right\}$ and $p = \left\{t,d,s \right\}$ quarks. The mixing parameters
$\delta_p$ are at least ${\cal O}(v/f)$ (their complete expressions are given in Ref.~\cite{Han:2005ru}), while $m_P$ are, naturally, ${\cal O}(f)$. 
\par
As it was pointed out before the SLH model has no custodial symmetry, i.e. there cannot be a $SU(2)_L \otimes SU(2)_R$ embedded into the $SU(2)_L \otimes U(1)_Y$ to which the $SU(3)_L \otimes U(1)_X$ breaks spontaneously.  However the $\rho \equiv M_W^2 /c_{\mbox{\tiny W}}^2 /M_Z^2 \simeq 1$ (or equivalently the $T$ oblique parameter) only gets corrections at ${\cal O}(v^2/f^2)$ and the breaking of the symmetry is very small. It is worthwhile to build a Little Higgs model 
preserving custodial symmetry and generating a collective Higgs quartic coupling free of quadratic divergences 
\cite{Schmaltz:2008vd}. A solution has been put forward in Ref.~\cite{Schmaltz:2010ac}.

\section{Lepton Flavour Violating Hadron Decays of the Tau Lepton}
The study of LFV in the SLH model has been carried out previously in Ref.~\cite{delAguila:2011wk} where $\mu \rightarrow e \gamma$, $\mu \rightarrow e e \overline{e}$
and $\mu -e$ conversion in nuclei where considered. Here we intend to apply the model for the study of LFV tau decays into hadrons, namely $\tau \rightarrow \mu P$, 
$\tau \rightarrow \mu V$, $\tau \rightarrow \mu PP$ where $P$ ($V$) is short for a pseudoscalar (vector) meson, that are of interest for Belle II and future flavour factories. 
\par
The procedure goes as follows. We have two different scales in the model: the vacuum expectation value of the SM Higgs, $v$, and the vacuum expectation value
of the triplets under $SU(3)$, $f$. Evidently we expect $v \ll f$ and in the limit $f \rightarrow \infty$ the effects of LFV should reduce to the negligible
ones of the SM (commented in the Introduction). Therefore we organize the calculation of the LFV amplitudes of the widths as an expansion in $v/f$ and we keep just the leading ${\cal O}(v^2/f^2)$ result. Our goal is to determine the amplitudes of the $\tau \rightarrow \mu q \overline{q}$ where $q= u,d,s$ quarks and, afterwards,
proceed to hadronize the corresponding quark bilinears. For this latter  step we will employ the tools given by chiral symmetry.

\subsection{$\tau \rightarrow \mu q \overline{q}$}
LFV decays in the SLH model arise at one-loop level and they are driven by the presence of the \lq \lq little" heavy neutrinos $N_i$ in connivance 
with the rotation of light lepton fields $V_{\ell}^{ij}$. There are two generic topologies participating in this amplitude: i) penguin-like diagrams, namely 
$\tau \rightarrow \mu \left\{\gamma, Z, Z' \right\}$, followed by $\left\{ \gamma, Z, Z' \right\} \rightarrow q \overline{q}$ and ii) box diagrams.
The calculation is obviously finite at this leading order. In principle there should be also a penguin-like contribution with a Higgs boson, i.e. 
$\tau \rightarrow \mu H$. However the coupling of the Higgs to the light quarks, $H \rightarrow q \overline{q}$,  has an intrinsic suppression due to the
 mass of the quarks and, therefore, we do not take this into account. In fact we will assume that light quarks are massless along all our calculation, and we will
 also neglect the muon mass. It has to be mentioned, however, that in Ref.~\cite{Celis:2013xja} it was pointed out that a one-loop Higgs generated gluon operator
 does not suffer of the light-quark mass suppression and could give a sizeable contribution. This would be independent of the LFV model employed. Although
 we are interested in the signatures specific to the SLH model and we do not include that gluon contribution in this article, 
 we think that this would require a separate analysis following up on our work here.
 \par
 Hence the full amplitude will be given by the sum of all contributions:
\begin{equation}  \label{eq:full}
{\cal T} = {\cal T}_{\gamma} + {\cal T}_Z + {\cal T}_{Z'} + {\cal T}_B \, . 
\end{equation}
\par 
We will use the unitary gauge. As it is well known, the number of Feynman diagrams is much reduced in this gauge because the only fields participating
in the dynamics are the physical ones. The price to pay is that the cancellation of divergences becomes rather intricate. While in the 't Hooft-Feynman gauge,
for instance, penguin and box diagrams are separately finite, in the unitary gauge they are not and the physical result is postponed until the final addition 
of all contributions. 
\par
Along the calculation we do a consistent expansion on the squared transfer momenta, i.e. $Q^2 = (p_{q} + p_{\overline{q}})^2$ over both the squared masses proportional to the $f$ scale ($M_{W'}$, $M_{Z'}$, $M_{N_i}$, $m_{P}$) and the SM gauge bosons. We only keep the leading order in this expansion. This amounts to an expansion, at the largest, in the $m_{\tau}^2/M_Z^2$ ratio. 

\par 
The diagrams contributing to the photon penguin are those in Figure~\ref{fig:1} and the result is given by:
%
\begin{figure}[t]
\centerline{\includegraphics[width=\linewidth]{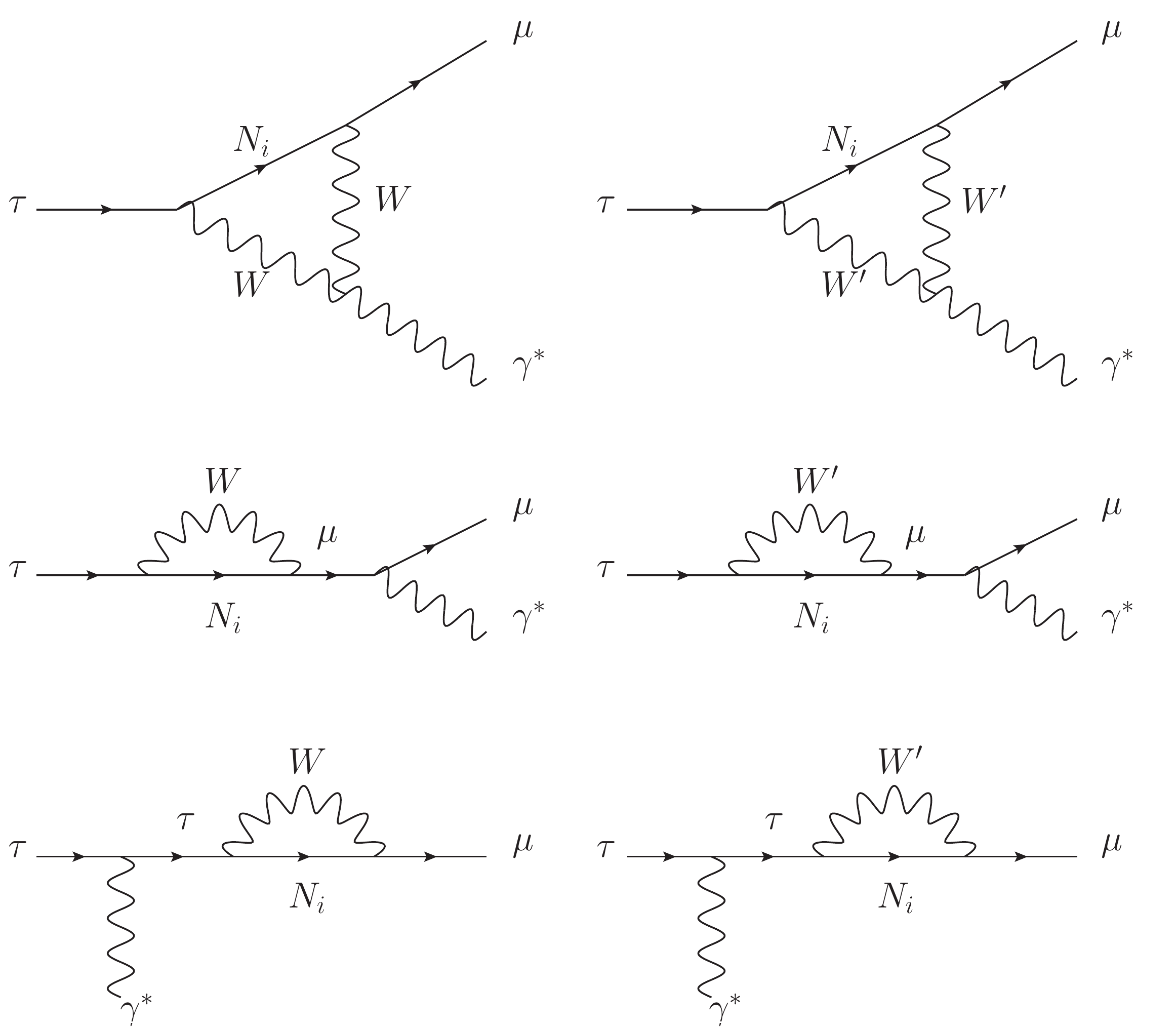}}
\caption{\label{fig:1} Penguin-like diagrams for $\tau \rightarrow \mu \gamma^*$ in the SLH model.}
\end{figure}
%
\begin{eqnarray}\label{eq:peng}
{\cal T}_ {\gamma} & = & \frac{e^2}{Q^2} \, \frac{v^2}{f^2} \,  \sum_j V_{\ell}^{j \mu *} V_{\ell}^{j \tau} \,  \overline{\mu}(p') \left[  Q^2 \gamma_\lambda \left( F_L^j P_L + F_R^j P_R \right)  \,
\right. \nonumber \\ 
& & +  \left.  i m_{\tau} \sigma_{\lambda \nu} Q^{\nu} \left( G_L^j P_L + G_R^j P_R \right) \right]  \tau(p) \,  \nonumber \\
& & \times \, \overline{q}(p_q) Q_q \gamma^\lambda q(p_{\overline{q}}) \, , 
\end{eqnarray}
where $P_{^L _R} = \left( 1 \mp \gamma_{5} \right)/2$ and $Q_q$ is the electric charge matrix: 
\begin{equation} \label{eq:qq}
Q_q =  \frac{1}{3} \left( \begin{array}{ccc}
               2& & \\
               & -1 & \\
               & & -1
               \end{array} \right),  
\end{equation} 
in units of $|e|$, and $q = \left( u, d, s  \right)^T$. With our assumptions above we find $F_R^j = {\cal O}(m_{\tau}^2 / M_{\mbox{\tiny{Z}}}^2)$, $G_L^j = 0$ and: 
\begin{eqnarray} \label{eq:fgs}
\! F_L^j  & = &   \frac{\alpha_{\mbox{\tiny W}}}{4 \pi} \frac{1}{16 M_{\mbox{\tiny W}}^2}  \left[ \left( \frac{\chi_j^3 \left( \chi_j^2 - 8 \chi_j + 13 \right)}{
\left( \chi_j - 1 \right)^4 } -4 \, \delta_{\nu}^2 \frac{M_{\mbox{\tiny W'}}^2}{M_W^2} \right) \, \ln \chi_j \right. \, \nonumber \\
&&  + \left. \frac{4 \chi_j^5 - 19 \chi_j^4 + 29 \chi_j^3 + 5 \chi_j^2 - 95 \chi_j + 40}{6 \left( \chi_j -1 \right)^3} \right] \! , \nonumber \\
\! G_R^j & = & \frac{\alpha_{\mbox{\tiny W}}}{4 \pi} \frac{1}{8 M_{\mbox{\tiny W}}^2}  \left[  \left( \frac{\chi_j^3 \left( 2 \chi_j + 1 \right)}{( \chi_j - 1)^4} + 2 \,  \delta_{\nu}^2 \frac{M_{\mbox{\tiny{W'}}}^2}{M_{\mbox{\tiny{W}}}^2} \right) \, 
\ln \chi_j \right.  \nonumber \\
& & + \left. \frac{6 \chi_j^5 - 15 \chi_j^4 - 35 \chi_j^3 + 72 \chi_j^2 - 66 \chi_j + 20}{6 (\chi_j - 1)^3} \right] , 
\end{eqnarray}
where $\alpha_{\mbox{\tiny W}} \equiv \alpha/s_{\mbox{\tiny W}}^2$ and $\chi_j = M_{N_j}^2/M_{\mbox{\tiny W'}}^2$. Notice that we have extracted a 
factor $v^2/f^2$ in Eq.~(\ref{eq:peng}) for the definition of the form factors. 
\par
\begin{figure}[t]
\centerline{\includegraphics[width=\linewidth]{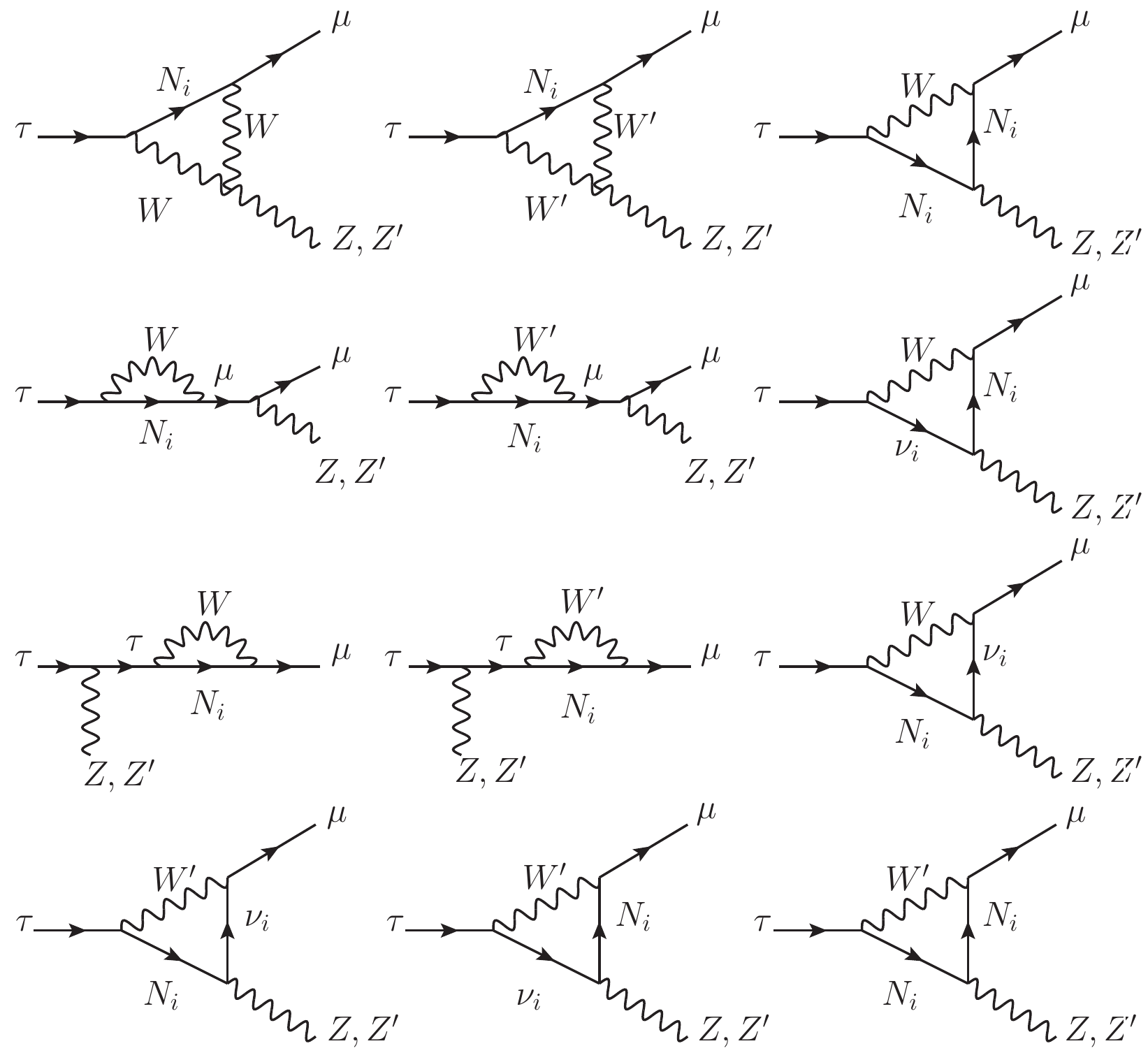}}
\caption{\label{fig:2} Penguin-like diagrams for $\tau \rightarrow \mu Z, Z'$ in the SLH model.}
\end{figure}

The penguin-like diagrams
with $Z$ and $Z'$ are given in Figure~\ref{fig:2}.
%
%
They give the following results:
\begin{eqnarray} \label{eq:fgz}
{\cal T}_Z & = & \frac{g}{M_{\mbox{\tiny Z}}^2} \, \sum_j V_{\ell}^{j \mu *} V_{\ell}^{j \tau} \, \overline{\mu}(p') \left[ \gamma_{\mu} \left( H_L^j P_L + H_R^j P_R \right) \right] \tau(p)  \nonumber \\
& & \times  \overline{q}(p_q) \left[ \gamma^{\mu} \left( Z_L P_L + Z_R P_R \right) \right] q(p_{\overline{q}}) , \nonumber \\
{\cal T}_{Z'} & = & \frac{g}{M_{\mbox{\tiny Z'}}^2} \, \sum_j V_{\ell}^{j \mu *} V_{\ell}^{j \tau} \, \overline{\mu}(p') \left[ \gamma_{\mu} \left( \widetilde{H}_L^{j} P_L + \widetilde{H}_R^{j}   P_R \right) \right] \tau(p)  \nonumber \\
& & \times \, \overline{q}(p_q) \left[ \gamma^{\mu} \left( Z_L' P_L + Z_R' P_R \right) \right] q(p_{\overline{q}}) , 
\end{eqnarray}
where now:
\begin{eqnarray} \label{eq:zqs}
Z_L &=& \frac{g}{c_{\mbox{\tiny{W}}}} \left( T_3^q - s_{\mbox{\tiny{W}}}^2 Q_q \right) , \nonumber \\
Z_R &=&  - \frac{g}{c_{\mbox{\tiny{W}}}} \,  s_{\mbox{\tiny{W}}}^2 \,  Q_q \, , \nonumber \\  
Z_L' &=& \frac{g}{6} \sqrt{3-t_{\mbox{\tiny W}}^2} \, \mathbb{1}_{\scriptscriptstyle 3 \times 3}\, , \nonumber \\
Z_R' &=& - \frac{g t_{\mbox{\tiny{W}}}^2}{\sqrt{3-t_{\mbox{\tiny W}}^2}} \, Q_q \, . 
\end{eqnarray}
being:
\begin{equation} \label{eq:t3q}
T_3^q = \frac{1}{2} \, \left( \begin{array}{ccc}
                              1 & & \\
                              & -1 & \\
                              & & -1 
                              \end{array}
                              \right) . 
\end{equation}
$H_R^j$ and $\widetilde{H}_R^j$ in Eq.~(\ref{eq:fgz}) are, again, ${\cal O}(m_{\tau}^2/M_{\mbox{\tiny{Z}}}^2)$ and we disregard them. For the 
\lq \lq left-handed" form factors we find: 
\begin{eqnarray} \label{eq:hfgz}
H_L^j & = & \frac{\alpha_{\mbox{\tiny{W}}}}{32 \pi} \left\{ \frac{\delta_{Z}}{c_{\mbox{\tiny{W}}}^2 \sqrt{3-t_{\mbox{\tiny{W}}}^2}} \right. \times  \nonumber \\
& & \left[
 \left( 3 \chi_j ( \chi_j - 2)  
 - 2 c_{\mbox{\tiny{W}}}^2  ( 7 \chi_j^2 - 14 \chi_j + 4 ) \right) \frac{\chi_j \, \ln \chi_j}{ (\chi_j-1)^2} \right. \nonumber \\
 & & \left.  + \frac{-5 \chi_j^2 + 5 \chi_j + 6 + 6 c_{\mbox{\tiny W}}^2 \left( 3 \chi_j^2 - \chi_j - 4 \right)}{2 ( \chi_j-1)} \right] \nonumber \\
 & & \left. - \delta_{\nu}^2 \, \frac{2 \chi_j^2 - 5 \chi_j+3}{ c_{\mbox{\tiny{W}}} ( \chi_j-1 )} \right\}  \, , \nonumber \\
\widetilde{H}_L^j & = & \frac{\alpha_{\mbox{\tiny{W}}}}{32 \pi}  \frac{1}{c_{\mbox{\tiny{W}}}^2 \sqrt{3-t_{\mbox{\tiny{W}}}^2}} \times  \nonumber \\
& & \left[
 \left( 3 \chi_j ( \chi_j - 2)  
 - 2 c_{\mbox{\tiny{W}}}^2  ( 7 \chi_j^2 - 14 \chi_j + 4 ) \right) \frac{\chi_j \, \ln \chi_j}{ (\chi_j-1)^2} \right. \nonumber \\
 & & \left.  + \frac{-5 \chi_j^2 + 5 \chi_j + 6 + 6 c_{\mbox{\tiny W}}^2 \left( 3 \chi_j^2 - \chi_j - 4 \right)}{2 ( \chi_j-1)} \right]  .
\end{eqnarray}
Notice that the result for $\widetilde{H}_L^j$ corresponding to the penguin-like $Z'$ contribution, and that is very similar to the result for
$H_L^j$, is ${\cal O}(1)$ in the $v/f$ expansion that we are performing. This is due to the fact that the definitions  of $H_L^j$ and $\widetilde{H}_L^j$ in 
Eq.~(\ref{eq:fgz}) carry a factor of the inverse squared mass of the reciprocal gauge boson in the penguin. Then the ${\cal T}_{\mbox{\tiny{Z'}}}$ amplitude
conveys the leading suppression factor in this factor term (see $M_{\mbox{\tiny{Z'}}}$ in Eq.~(\ref{eq:gmass})). 
\par
\begin{figure}[t]
\centerline{\includegraphics[width=0.8\linewidth]{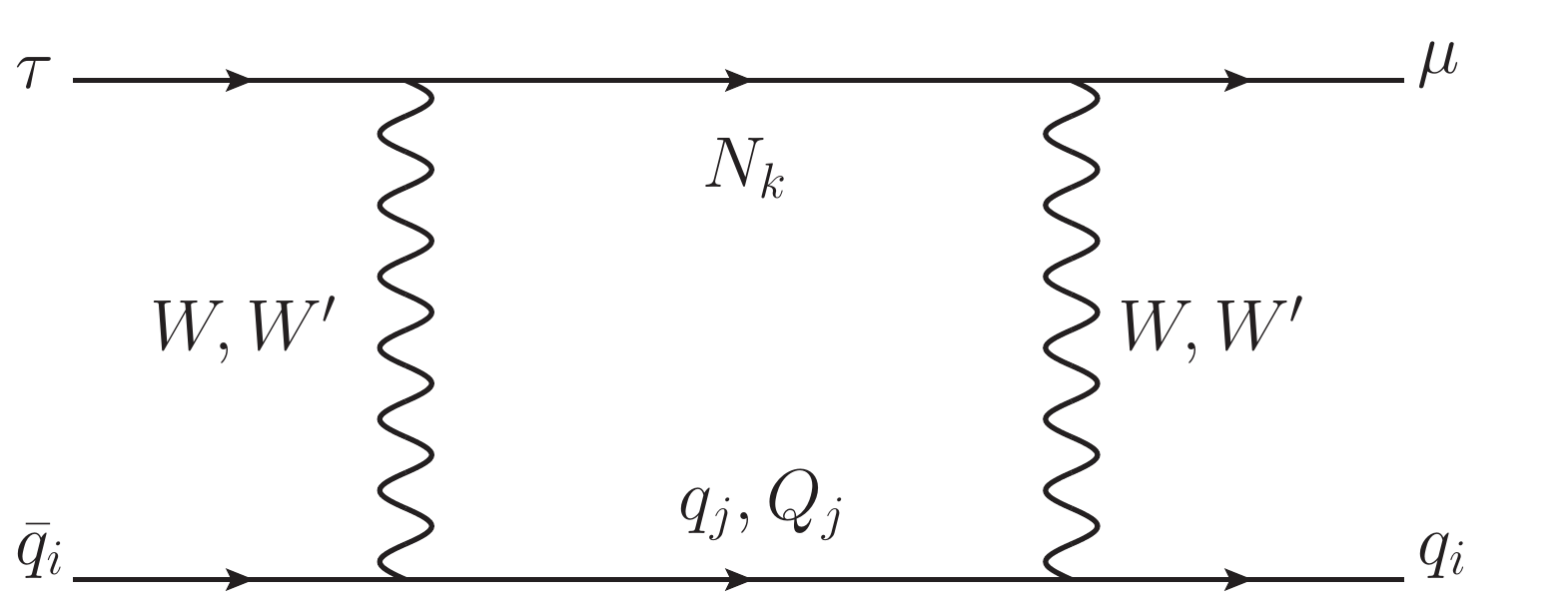}}
\caption{\label{fig:3} Box diagrams for $\tau \rightarrow \mu q \overline{q}$ in the SLH model. The internal quark states are $(u,\overline{u}) \rightarrow 
\{d, D\}$, 
$(d, \overline{d}) \rightarrow \{u \}$, $(s, \overline{s}) \rightarrow \{c\}$.} 
\end{figure}

 
Finally we turn to evaluate the box diagrams in Figure~\ref{fig:3}. We proceed following the same approaches as in the case of the penguin diagrams. In 
addition we consider that the external momenta vanish.
%
%
The result is given by:
\begin{eqnarray} \label{eq:boxee}
{\cal T}_B &=& g^2 \, \sum_{q}^{u,d,s} \, \sum_j V_{\ell}^{j \mu *} V_{\ell}^{j \tau} \; B_q^j \, \nonumber \\
& & \times  \, \overline{\mu}(p') \gamma_{\mu}  P_L  \tau(p) \, \cdot \,   \overline{\psi_q}(p_q)  \gamma^{\mu}  P_L   \psi_q(p_{\overline{q}}) , 
\end{eqnarray}
where $\psi_q = \left\{u,d,s\right\}$ and 
\begin{equation} \label{eq:bqj}
B_q^j \, = \, \frac{\alpha_{\mbox{\tiny{W}}}}{64 \pi} \, \left[\alpha_q^j \, \ln \chi_j \, + \, \beta_q^j \ln \delta \, + \, \gamma_q^j \right] \, ,
\end{equation}
and $\delta = m_D^2/M_{\mbox{\tiny{W'}}}^2$. The remaining terms are given by:
\begin{eqnarray} \label{eq:boxfac}
 \alpha_u^j &=& \frac{1}{M_{\mbox{\tiny{W}}}^2 (\chi_j-\delta)}  \left\{ \frac{3 \chi_j \delta \delta_{\nu} ( \delta_d^* + \delta_d )}{( \chi_j-1)}  \right. 
 - \delta^2 \delta_{\nu} ( \delta_d^* + \delta_d ) \, \nonumber \\
 & & \left. + \frac{\chi_j ( 6 - 13 \chi_j )}{(\chi_j-1)^2} \frac{M_{\mbox{\tiny{W}}}^2}{M_{\mbox{\tiny{W'}}}^2 } 
 + \, (\delta^2 - 6 \delta) \frac{M_{\mbox{\tiny{W}}}^2}{M_{\mbox{\tiny{W'}}}^2} \, \right. \nonumber \\
 & & \left. + \delta^2 \delta_d^2 \delta_{\nu}^2  \frac{M_{\mbox{\tiny{W'}}}^2}{M_{\mbox{\tiny{W}}}^2} \right\} ,  \nonumber \\
 \alpha_d^j & = & \frac{3 \, \delta_{\nu}}{M_{\mbox{\tiny{W}}}^2 (\chi_j-1)} \, \left( \delta_d^* + \delta_d \right) , \nonumber \\
 \alpha_s^j & = & \frac{3 \, \delta_{\nu}}{M_{\mbox{\tiny{W}}}^2 (\chi_j-1)} \, \left( \delta_s^* + \delta_s \right) , \nonumber \\ 
\beta_u^j & = & \frac{\delta^2}{M_{\mbox{\tiny{W}}}^2 (\delta - \chi_j)} \left\{ \delta_d^2 \delta_{\nu}^2 \frac{M_{\mbox{\tiny{W'}}}^2}{M_{\mbox{\tiny{W}}}^2}
+ \frac{\delta (\delta - 8)}{(\delta -1)^2} \frac{M_{\mbox{\tiny{W}}}^2}{M_{\mbox{\tiny{W'}}}^2} \right. 	\nonumber \\
& & \left. - \delta_{\nu} (\delta_d^* + \delta_d) \frac{\delta^2 - 5 \delta + 4}{(\delta -1)^2} \right\} ,\, \nonumber \\
\beta_d^j & = & \beta_s^j \, = 0 \,,  \nonumber \\
\gamma_u^j & = & - \, \frac{1}{2 M_{\mbox{\tiny{W}}}^2} \left\{ 3 \, \delta_{\nu}^2 \, \delta_d^2 \, \chi_j \frac{M_{\mbox{\tiny{W'}}}^2}{M_{\mbox{\tiny{W}}}^2} \right.
\nonumber \\
& &  \left. + \frac{ \delta (3 \chi_j^2 -16 \chi_j + 13) - 3 \chi_j^2 + 13 \chi_j + 4}{(\delta -1 ) (\chi_j-1)} 
\frac{M_{\mbox{\tiny{W}}}^2}{M_{\mbox{\tiny{W'}}}^2} \right\} , \nonumber \\
\gamma_d^j & = & \frac{3}{2 M_{\mbox{\tiny{W}}}^2} \, \delta_{\nu} \, (\delta_d^* + \delta_d) \, \chi_j ,\, \nonumber \\
\gamma_s^j & = & \frac{3}{2 M_{\mbox{\tiny{W}}}^2} \, \delta_{\nu} \, (\delta_s^* + \delta_s) \, \chi_j \, . 
\end{eqnarray}
Here the $\delta_d$ and $\delta_s$ parameters have been defined at the end of Section II. 

\subsection{Hadronization}
Our results for the full amplitude ${\cal T}$ in Eq.~(\ref{eq:full}) are given in terms of light quark bilinears with different weights provided by the 
theory. As our goal is to study the final states of one meson (either pseudoscalar o vector) and two pseudoscalar mesons, we need to implement a procedure
in order to hadronize the quark bilinears. An essentially model-independent scheme is the one provided by Chiral Perturbation Theory
\cite{Gasser:1983yg,Gasser:1984gg,Ecker:1988te}. We follow the procedure and definitions put forward in Ref.~\cite{Arganda:2008jj} where all the 
expressions are fully given.  
\par 
The quark bilinears that appear in ${\cal T}$ can be written in terms of the QCD quark currents:
\begin{equation} \label{eq:qcdcur}
V_{\mu}^i \, = \, \overline{q} \, \gamma_{\mu} \,  \frac{\lambda^i}{2} \, q \, \qquad , \qquad 
A_{\mu}^i \, = \, \overline{q} \,  \gamma_{\mu}  \, \gamma_5 \, \frac{\lambda^i}{2} \,  q \, ,
\end{equation}
where we remind that $q = (u,d,s)^T$. For instance, the electromagnetic quark current in Eq.~(\ref{eq:peng}) reads:
\begin{equation} \label{eq:emc}
{\cal J}_{\mu}^{\mbox{\tiny{em}}} = \overline{q} \, Q_q \, \gamma_{\mu} q \, = \, V_{\mu}^3 \, + \, \frac{1}{\sqrt{3}} \, V_{\mu}^8 \, , 
\end{equation}
and
\begin{eqnarray}\label{eq:vmac}
\overline{u} \, \gamma_{\mu} \, P_L \, u & = & J_{\mu}^3 + \frac{1}{\sqrt{3}} V_{\mu}^8 + \frac{2}{\sqrt{6}} J_{\mu}^0 \, , \nonumber \\
\overline{d} \, \gamma_{\mu} \, P_L \, d & = & - \, J_{\mu}^3 + \frac{1}{\sqrt{3}} V_{\mu}^8 + \frac{2}{\sqrt{6}} J_{\mu}^0 \, , \nonumber \\
\overline{s} \, \gamma_{\mu} \, P_L \, s & = & -  \frac{2}{\sqrt{3}} V_{\mu}^8 + \frac{2}{\sqrt{6}} J_{\mu}^0 \, ,
\end{eqnarray} 
with $J_{\mu}^i = (V_{\mu}^i - A_{\mu}^i)/2$. The QCD currents are determined as the functional derivatives, with respect to the external 
auxiliary fields, of the Resonance Chiral Theory action ${\cal L}_{\mbox{\tiny{R$\chi$T}}}$ \cite{Ecker:1988te}:
\begin{equation} \label{eq:deric}
V_{\mu}^i = \frac{\partial {\cal L}_{\mbox{\tiny{R$\chi$T}}}}{\partial v^{\mu}_i} \, \Bigg|_{j=0}   \, ,  \qquad  \; \; \; 
A_{\mu}^i = \frac{\partial {\cal L}_{\mbox{\tiny{R$\chi$T}}}}{\partial a^{\mu}_i} \, \Bigg|_{j=0} \,   
\end{equation}
where $j=0$ indicates that, after derivation, all the external currents are put to zero. The vector current contributes to an even number of pseudoscalar
mesons or a vector resonance, while the axial-vector current gives an odd number of pseudoscalar mesons. 

\subsubsection{$\tau \rightarrow \mu P$}
Only the axial-vector current contributes and that means that ${\cal T}_{\gamma}$ does not participate. The axial-vector current is determined from 
the leading ${\cal O}(p^2)$ chiral Lagrangian and we get, for $P =\left\{ \pi^0, \eta, \eta' \right\}$:
\begin{eqnarray} \label{eq:unop}
{\cal T}_{Z}(P) & = & - i \frac{g^2}{2 c_{\mbox{\tiny{W}}}} \frac{F}{M_{\mbox{\tiny{Z}}}^2}  \, Z(P) \sum_j V_{\ell}^{j \mu *} V_{\ell}^{j \tau} \nonumber \\
& & \times \, \overline{\mu}(p') \left[ \slashed{Q} \left(H_L^j P_L + H_R^j P_R \right) \right] \tau(p), \nonumber \\
{\cal T}_{Z'}(P) & = &  i \frac{g^2}{4 \sqrt{9-3 t_{\mbox{\tiny{W}}}^2}} \frac{F}{M_{\mbox{\tiny{Z'}}}^2}  \, Z'(P) \sum_j V_{\ell}^{j \mu *} V_{\ell}^{j \tau} \nonumber \\
& & \times \, \overline{\mu}(p') \left[ \slashed{Q} \left(\widetilde{H}_L^j P_L + \widetilde{H}_R^j P_R \right) \right] \tau(p), \nonumber \\
{\cal T}_B(P) & = & -i g^2 F \sum_j V_{\ell}^{j \mu *} V_{\ell}^{j \tau} \,  \nonumber \\
& &  \times \, B^j(P) \; \overline{\mu}(p') \left[ \slashed{Q} P_L \right] \tau(p) .
\end{eqnarray}  
Here $F \simeq 0.0922 \, \mbox{GeV}$ is the decay constant of the pion and the $Z(P)$, $Z'(P)$ and $B_j(P)$ factors are given in Table~\ref{tab:1}. 
\par 
\begin{table*}
\setlength{\tabcolsep}{10pt}
\begin{tabular}{|c|c|c|c|} \hline
& & & \\ 
& $P=\pi^0$ & $P=\eta$ & $P=\eta'$ \\ 
& & & \\ \hline
& & & \\
$Z(P)$ & $1$ & $\frac{1}{\sqrt{6}} \, \left( \sin \theta_{\eta} + \sqrt{2} \cos \theta_{\eta} \right)$ &
$ \frac{1}{\sqrt{6}} \, \left( \sqrt{2} \sin \theta_{\eta} - \cos \theta_{\eta} \right)$ \\ 
& & & \\ \hline
& & & \\
$Z'(P)$ & $\sqrt{3} t_{\mbox{\tiny{W}}}^2$  & $\cos \theta_{\eta} t_{\mbox{\tiny{W}}}^2 - \sqrt{2} \sin \theta_{\eta} \left(3 - t_{\mbox{\tiny{W}}}^2 \right)$ &
$\sin \theta_{\eta} t_{\mbox{\tiny{W}}}^2 + \sqrt{2} \cos \theta_{\eta} \left(3 - t_{\mbox{\tiny{W}}}^2 \right)$ \\ 
& & & \\  \hline
& & & \\
$B_j(P)$ & $\frac{1}{2} \left( B_d^j - B_u^j \right) $ & $\frac{1}{2 \sqrt{3}} \left[ \left( \sqrt{2} \sin \theta_{\eta} - \cos \theta_{\eta} \right) B_u^j  \right.$
& $\frac{1}{2 \sqrt{3}} \left[ \left( \sin \theta_{\eta} - 2 \sqrt{2} \cos \theta_{\eta} \right) B_d^j \right.$ \\ & & $+ \left.  \left( 2 \sqrt{2} \sin \theta_{\eta} + \cos \theta_{\eta} \right) B_d^j \right] $ & 
$ - \left. \left( \sin \theta_{\eta} + \sqrt{2} \cos \theta_{\eta} \right) B_u^j \right]$  \\
& & & \\ \hline
\end{tabular}
\caption{Factors appearing in Eq.~(\ref{eq:unop}). The mixing between the octet ($\eta_8$) and the singlet ($\eta_0$) components of the nonet of pseudoscalar mesons is parameterized by the angle $\theta_{\eta} \simeq - 18^{\circ}$. The functions $B_q^j$ are given in Eq.~(\ref{eq:bqj}).}
\label{tab:1}
\end{table*}
The width of these processes, with ${\cal T}(P) = {\cal T}_{Z}(P) + {\cal T}_{Z'}(P) + {\cal T}_{B}(P)$, is given by:
\begin{equation} \label{eq:tup}
B(\tau \rightarrow \mu P) =  \frac{\lambda^{1/2}(m_{\tau}^2,m_{\mu}^2,m_P^2)}{4 \pi \, m_{\tau}^2 \, \Gamma_{\tau}} \, \frac{1}{2} \sum_{i,f} \left| {\cal T}(P) \right|^2 \, ,
\end{equation}
where $\lambda(x,y,z) = (x+y-z)^2-4xy$, and
\begin{eqnarray} \label{eq:abes}
 \sum_{i,f} \left| {\cal T}(P) \right|^2 &=& \frac{1}{2 m_{\tau}} \sum_{k,l} \left[(m_{\tau}^2+m_{\mu}^2-m_P^2) \left( a_P^k a_P^{l*} + b_P^k b_P^{l*} \right)
 \right. \nonumber \\
& & \left. + 2 m_{\mu} m_{\tau} \left( a_P^k a_P^{l*} - 
b_P^k b_P^{l*} \right)   \right] \, , 
\end{eqnarray}
with $k,l = Z, Z', B$. Defining $\Delta_{\tau \mu} = m_{\tau}- m_{\mu}$, $\Sigma_{\tau \mu} = m_{\tau}+ m_{\mu}$ we have :
\begin{eqnarray} \label{eq:abesex}
a_P^Z & = & - \frac{g^2 \, F}{4 c_{\mbox{\tiny{W}}} M_{\mbox{\tiny{Z}}}^2} \, \Delta_{\tau \mu}\,  Z(P) \sum_j V_{\ell}^{j \mu *} V_{\ell}^{j \tau} \left(H_R^j + H_L^j
\right) , \nonumber \\
a_P^{Z'} & = & \frac{g^2 \, F}{8 \sqrt{9-3 t_{\mbox{\tiny{W}}}^2} M_{\mbox{\tiny{Z'}}}^2} \, \Delta_{\tau \mu} \,  Z'(P) \sum_j V_{\ell}^{j \mu *} V_{\ell}^{j \tau} \left(\widetilde{H}_R^j + \widetilde{H}_L^j
\right) , \nonumber \\
a_P^B &=& -  \frac{g^2 \, F}{2} \, \Delta_{\tau \mu} \,  \sum_j V_{\ell}^{j \mu *} V_{\ell}^{j \tau} B_j(P) ,\nonumber \\
b_P^Z &=&  \frac{g^2 \, F}{4 c_{\mbox{\tiny{W}}} M_{\mbox{\tiny{Z}}}^2} \, \Sigma_{\tau \mu} \,  Z(P) \sum_j V_{\ell}^{j \mu *} V_{\ell}^{j \tau} \left(H_R^j - H_L^j
\right) , \nonumber \\
b_P^{Z'} &=& - \frac{g^2 \, F}{8 \sqrt{9-3 t_{\mbox{\tiny{W}}}^2} M_{\mbox{\tiny{Z'}}}^2} \, \Sigma_{\tau \mu} \,  Z'(P) \sum_j V_{\ell}^{j \mu *} V_{\ell}^{j \tau} \left(\widetilde{H}_R^j - \widetilde{H}_L^j
\right) , \nonumber \\
b_P^B &=& -  \frac{g^2 \, F}{2} \, \Sigma_{\tau \mu} \,  \sum_j V_{\ell}^{j \mu *} V_{\ell}^{j \tau} B_j(P) . 
\end{eqnarray}

\subsubsection{$\tau \rightarrow \mu P P$}
Here we will consider the decays into the pairs $P \overline{P} = \left\{ \pi^+ \pi^-, K^+ K^-, K^0 \overline{K^0} \right\}$. 
To the final state of two pseudoscalar mesons all vector components of the different pieces discussed above, i.e. $\gamma, Z, Z'$-penguins and the box amplitude, 
contribute. The hadronization is driven by the vector form factor that can be defined through the electromagnetic current in Eq.~(\ref{eq:emc}):
\begin{equation} \label{eq:fv}
\langle P_1(p_1) P_2(p_2) | {\cal J}_{\mu}^{\mbox{\tiny{em}}} | 0 \rangle \, = \, (p_1 - p_2)_{\mu} \, F_V^{P_1 P_2}(Q^2) , 
\end{equation}
where $Q=p_1+p_2$. The determination of the vector form factor has a long story and we will not dwell on it here. We will consider its construction in the 
frame of the chiral theory and will take the results put forward in Ref.~\cite{Arganda:2008jj}, except for the pion case where we take the 
improved version of Refs.~\cite{Shekhovtsova:2012ra,Dumm:2013zh}.
\par 
After hadronization we obtain:
\begin{eqnarray} \label{eq:hvector}
{\cal T}_{\gamma}^P &=&  \frac{e^2}{Q^2} \, \frac{v^2}{f^2} \,  F_V^{P \overline{P}}(Q^2) \times \nonumber \\
&&  \sum_j  V_{\ell}^{j \mu *} V_{\ell}^{j \tau} \,  \overline{\mu}(p') \left[  Q^2 
(\slashed{p}_q - \slashed{p}_{\overline{q}}) \left( F_L^j P_L + F_R^j P_R \right) \right. \nonumber \\ 
& & +  \left. 2 i m_{\tau} p_q^{\lambda} \sigma_{\lambda \nu} p_{\overline{q}}^{\nu} \left( G_L^j P_L + G_R^j P_R \right) \right]  \tau(p) \, ,  \nonumber \\
{\cal T}_Z^P & = & g^2 \frac{2 s_{\mbox{\tiny{W}}}^2-1}{2 c_{\mbox{\tiny{W}}} M_{\mbox{\tiny{Z}}}^2} \,  F_V^{P \overline{P}}(Q^2) \times \nonumber \\
& &   \sum_j  V_{\ell}^{j \mu *} V_{\ell}^{j \tau} \, \overline{\mu}(p') (\slashed{p}_q - \slashed{p}_{\overline{q}}) \left( H_L^j P_L + H_R^j P_R \right) \, \tau(p) \, , \nonumber \\ 
{\cal T}_{Z'}^P & = & - g^2 \frac{t_{\mbox{\tiny{W}}}^2}{4  M_{\mbox{\tiny{Z'}}}^2 \sqrt{3 -t_{\mbox{\tiny{W}}}^2} } \,  F_V^{P \overline{P}}(Q^2) 
\times  \nonumber \\
& &  \sum_j  V_{\ell}^{j \mu *} V_{\ell}^{j \tau} \, \overline{\mu}(p') (\slashed{p}_q - \slashed{p}_{\overline{q}}) \left( \widetilde{H}_L^j P_L + \widetilde{H}_R^j P_R \right) \, \tau(p)  \, , \nonumber \\ 
{\cal T}_B^P & = & \frac{g^2}{2} \, F_V^{P \overline{P}}(Q^2) \, \times \nonumber \\
& &  \sum_j  V_{\ell}^{j \mu *} V_{\ell}^{j \tau} \, \left(B_u^j - B_d^j \right)  \overline{\mu}(p') (\slashed{p}_q - \slashed{p}_{\overline{q}})  P_L \tau(p) . 
\end{eqnarray}
\par
The branching ratio for the process is:
\begin{equation} \label{eq:brtmpp}
B(\tau \rightarrow \mu PP) = \frac{\kappa}{64 \pi^3 m_{\tau}^2 \Gamma_{\tau}} \int_{s_-}^{s_+} ds \int_{t_-}^{t_+} dt \,  \frac{1}{2} \sum_{i,f} |{\cal T}^P|^2 ,
\end{equation}
where $\kappa=1$ for $\pi^+ \pi^-$, $K^+ K^-$ and $K^0 \overline{K^0}$. In Eq.~(\ref{eq:brtmpp}), and in terms of the momenta of the
particles participating in the process, $s = (p_q + p_{\overline{q}})^2$ and $t = (p-p_{\overline{q}})^2$. Moreover  ${\cal T}^P = {\cal T}_{\gamma}^P + 
{\cal T}_{Z}^P + {\cal T}_{Z'}^P + {\cal T}_B^P$, and the integration limits are:
\begin{eqnarray} \label{eq:lint}
t_-^+ & = &   \frac{1}{4s} \left[ (m_{\tau}^2-m_{\mu}^2)^2 - \left( \lambda^{1/2}(s,m_P^2,m_P^2)  \right. \right.  \nonumber \\
& & \left. \left. \qquad \mp \lambda^{1/2}(m_{\tau}^2,s,m_{\mu}^2)\right)^2 \right] \, , \nonumber \\
s_- & = & 4 m_P^2 \, , \nonumber \\
s_+ &=& (m_{\tau} - m_{\mu})^2 \, .  
\end{eqnarray}

\subsubsection{$\tau \rightarrow \mu V$}
We would like to consider also the decays into a vector resonance, namely $V = \rho, \phi$. From a quantum field theory point of view, a resonance is not
an asymptotic state and, indeed, a vector decays strongly into a pair of pseudoscalar mesons. When an experiment {\it measures} a final state with a vector resonance, in fact what is measuring is a pair of pseudoscalar mesons with a squared total mass approaching $m_V^2$. Hence the definition of a resonance from an experimental point of view is uncertain. Actually the chiral nature of the lightest pseudoscalar mesons realizes on this occurrence and two pions into a $J=I=1$ state are
indistinguishable from a $\rho(770)$ meson, for instance. As a consequence the channels $\tau \rightarrow \mu V$ are related with $\tau \rightarrow \mu PP$ that we discussed above. We follow the proposal in Ref.~\cite{Arganda:2008jj}. 
\par
The outcome of this circumstance is that the branching ratio of $\tau \rightarrow \mu V$ is obtained from that of the $\tau \rightarrow \mu PP$ 
by trying to implement the experimental procedure, that is, focusing in two pseudoscalar mesons on the mass (and width) of the resonance. That is:
\begin{eqnarray} \label{eq:tmrho}
B(\tau \rightarrow \mu \rho) & = &  B(\tau \rightarrow \mu \pi^+ \pi^-) \Big|_{\rho}  \, ,  \\
B(\tau \rightarrow \mu \phi) & = &  B(\tau \rightarrow \mu K^+ K^-) \Big|_{\phi} + B(\tau \rightarrow \mu K^0 \overline{K^0}) \Big|_{\phi} \, , \nonumber
\end{eqnarray}
where the two pseudoscalars branching ratio is the one given by Eq.~(\ref{eq:brtmpp}) but where the $s_{\pm}$ limits of integration are now specified
by:
\begin{equation} \label{eq:slimits1}
s_{\pm} \, = \,  M_{\rho}^2 \, \, \pm \,  \frac{1}{2} \, M_{\rho} \,  \Gamma_{\rho}(M_{\rho}^2) \, . 
\end{equation}
and
\begin{equation} \label{eq:slimits2}
s_{\pm} \, = \,  M_{\phi}^2 \, \, \pm \,  \frac{1}{2} \, M_{\phi} \,  \Gamma_{\phi}(M_{\phi}^2) \, ,
\end{equation}
respectively. Here the total widths of $\rho$ and $\phi$ are taken from Ref.~\cite{Agashe:2014kda}.  We think that this definition of the branching ratios into vector mesons approaches the experimental interpretation and provides a reasonable estimate of them. 

\section{Numerical Results}
The provision of numerical estimates for our LFV branching ratios, from our results in the previous Section, requires an all-inclusive discussion of the 
parameters of the SLH model that we have employed:
\par
- \underline{Scale of compositeness $f$}. As commented in the Introduction almost everyone expects some new physics around $E \sim 1 \, \mbox{TeV}$, and 
going up. We could fix the scale of compositeness in the SLH model as that $f \sim 1 \, \mbox{TeV}$. However analyses of the model from Higgs data and Electroweak
Precision Observables \cite{Reuter:2012sd,Reuter:2013zja} seem to indicate that, at $95 \, \%$ C.L., values of $f \lsim 3.5 \, \mbox{TeV}$ should be 
excluded for our model. That, of course, also delays the appearance of a strongly coupled region. For definiteness we choose a range 
$ 2 \, \mbox{TeV} < f < 10 \, \mbox{TeV}$ in order to furnish our results.
\par 
- \underline{Heavy neutrinos}. \lq \lq Little" neutrinos drive the dynamics of LFV lepton decays. Inherited from the SM setting we have three different
heavy neutrinos that appear in the amplitudes in Eqs.~(\ref{eq:peng},\ref{eq:fgz},\ref{eq:boxee}) and that provide an amplitude (\ref{eq:full}) that we
can write, generically, as:
\begin{equation} \label{eq:sumneu}
{\cal T} \, = \, \sum_{j} V_{\ell}^{j \mu *} V_{\ell}^{j \tau} \, A \left(\chi_j \right)  ,
\end{equation}
with $j$ adding over the three families and $A(\chi_j)$ a generic function of $\chi_j = M_{N_j}^2/M_{\mbox{\tiny{W'}}}^2$. We do not have any information on the mixing matrix elements $V_{\ell}^{ik}$ and
we have to keep at least two families in order to have a non-vanishing result; as a consequence we will give our numerical results assuming only two families
and, accordingly, one mixing angle. Hence we will have:
\begin{equation} \label{eq:sumneu2}
{\cal T} \, = \, \sin \theta \, \cos \theta \, \left[ A(\chi_1) - A(\chi_2) \right] \, . 
\end{equation}
In Ref.~\cite{delAguila:2011wk} it can be observed that, from LFV tau decays into leptons within the SLH model and for $f \simeq 1 \, \mbox{TeV}$, experimental
bounds require $\sin 2 \theta = 2 \sin \theta \cos \theta < 0.05$
and even smaller from muon-electron conversion in nuclei. As we propose higher values for
the scale of compositeness we will take, for the numerical determinations $\sin 2 \theta \simeq 0.25$, though we will also study
 the variation of the branching ratios in function of this parameter.  
 \par
\lq \lq Little" neutrino masses are also unknown. 
Experimental bounds on these masses are rather loose and very much model dependent \cite{Deppisch:2015qwa}.
However, we will take into account the results in Ref.~\cite{delAguila:2011wk} pointing to $\chi_1\chi_2\lesssim0.01$ and $\sqrt{\chi_1/\chi_2}-\sqrt{\chi_2/\chi_1}\lesssim 0.05$. Given our larger 
values for $f$ we will use ($\chi_2>\chi_1$ is assumed, our spectrum cannot be degenerated) $0\leq \chi_1\leq0.25$ and $1.1\chi_1\leq \chi_2\leq 10\chi_1$, where the latter limits of 
$\chi_2$ correspond to the nearly-degenerate and large mass-splitting cases, respectively. 
\par 
- \underline{$\tan \beta = f_1/f_2$}. The ratio of the two vevs from the spontaneous breaking of the upper symmetry is also an unknown parameter in our model. 
The mixing between a \lq \lq little" and a light neutrino, parameterized by $\delta_{\nu}$ in Eq.~(\ref{eq:deltanu}), can give us a hint. Phenomenological
analyses indicate that $\delta_{\nu} < 0.05$ \cite{delAguila:2011wk,deBlas:2013gla,Deppisch:2015qwa}. Therefore from Eq.~(\ref{eq:deltanu}) we obtain
that $|f \, t_{\beta}| \gsim 3.5 \, \mbox{TeV}$. We will take, as a value of reference, $t_{\beta} = 5$ and will explore the range $ 1 < t_{\beta} < 10$. 
\par 
- \underline{Quark parameters}. As commented above we do not consider flavour-mixing in the quark sector. The redefinition of fields that diagonalizes the mixing between \lq \lq little" and light quarks is parameterized
by the $\delta_p$ parameters that appear in the box amplitude, Eq.~(\ref{eq:boxfac}), for $p=d,s$. We follow the proposal of Ref.~\cite{Han:2005ru} and assume
that the mixing effects in the down-quark sector are suppressed in the $t_{\beta} > 1$ regime. This is analogous to what happens in the neutrino
case. It implies:
\begin{equation} \label{eq:deltasnu}
\delta_d \simeq \delta_s \simeq - \delta_{\nu} . 
\end{equation} 
Finally, in the box diagrams also appears the ratio $\delta = m_D^2/M_{\mbox{\tiny{W'}}}^2$ that involves the mass of the \lq \lq little" down quark D. In all
the numerical evaluations we take $\delta = 1$. 
\par 
The input of SM parameters and masses is taken from the PDG \cite{Agashe:2014kda}. In particular we will take $\sin^2 \theta_{\mbox{\tiny{W}}} = 0.23$, 
$F = 0.0922 \, \mbox{GeV}$ and $\theta_{\eta} = - 18^{\circ}$. 
\par 
Although we will present our results for LFV tau decays into a muon and hadrons, the results on the decay to an electron
should be essentially the same because we have expanded the mass of the outgoing charged lepton over heavy masses in our 
calculation. Hence the only difference between both channels is, essentially, one of phase space that would turn out to be
tiny in any case due to the relative high mass of the tau lepton. Hence we consider our results to be valid for both decays: $\tau \rightarrow \ell \, \mbox{hadrons}$ for $\ell = e, \mu$.
\begin{table}
\setlength{\tabcolsep}{20pt}
\begin{tabular}{|c|c|c|} \hline
\raisebox{-1.8ex}{Process} & \multicolumn{2}{c|}{\raisebox{-0.4ex}{$B \times 10^{\mbox{\tiny{8}}}$ ($90 \, \% $ C.L.) \cite{Agashe:2014kda}}}  \\ 
 \cline{2-3}  
 & $\ell= \mu$ & $ \ell=e$ \\ \hline
$\tau \rightarrow \ell \, \gamma$ & $< 4.4$ & $< 3.3$\\ 
 \hline
$\tau \rightarrow \ell \, \pi^{\mbox{\tiny{0}}}$ & $< 11.0$ & $<8.0$\\
\hline
$\tau \rightarrow \ell \, \eta$ & $< 6.5 $ & $<9.2$\\
 \hline
$\tau \rightarrow \ell \, \eta'$ & $ < 13.0$ & $<16.0$\\
 \hline
$\tau \rightarrow \ell \, \pi^+ \pi^-$ & $<2.1$ & $<2.3$\\
 \hline
$\tau \rightarrow \ell \,  K^+ K^-$ & $<4.4$ & $<3.4$\\
 \hline
$\tau \rightarrow \ell \, K_{\mbox{\tiny{S}}} \overline{K}_{\mbox{\tiny{S}}}$ & $ < 8.0$ & $<7.1$\\
 \hline
$\tau \rightarrow \ell \, \rho^{\mbox{\tiny{0}}}$ & $< 1.2 $ & $<1.8$\\
 \hline
$\tau \rightarrow \ell \, \phi$ & $< 8.4 $ & $<3.1$\\
 \hline
\end{tabular}
\caption{Experimental upper bounds, at $90 \, \%$ C.L., on the branching ratios of the LFV decays $\tau \rightarrow \ell (P,V,PP)$ for $\ell = \mu,e$, studied in this article. We quote them from the PDG \cite{Agashe:2014kda}.}
\label{tab:2}
\end{table}
\par 
The present upper bounds on the LFV hadron tau decays branching ratios, that we study in this article, are collected in Table~\ref{tab:2}. All these bounds originate in the excellent work carried out by both BaBar and Belle experiments in the last ten years. It can be seen that present limits stand at the $10^{-8}$ level.
Super B Factories, like the SuperKEKB/Belle II project \cite{Shwartz:2015kja} will give the next step. Hadron decays of the tau lepton are almost
background free, although efficiencies are different from channel to channel. All in all, expected sensitivities are in the range of 
$B( \tau \rightarrow \ell \,  \mbox{hadrons} ) \sim (2-6) \times 10^{-10}$ \cite{CeiA:2014wea}. 
\par
\begin{figure}[!]
\centerline{\includegraphics[width=1.15\linewidth]{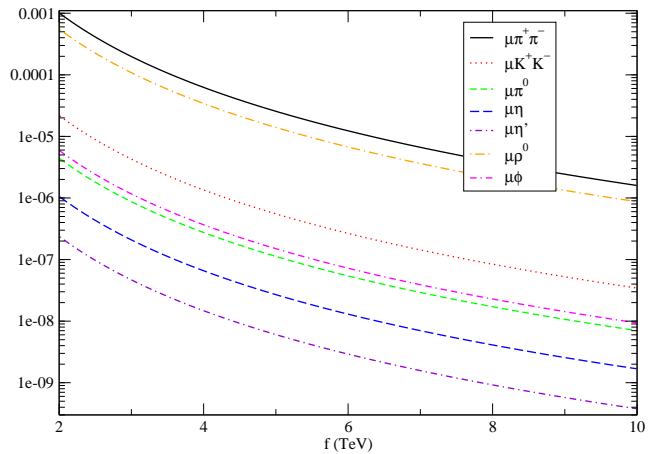}}
\caption{\label{fig:4} Dependence of the scale of compositeness $f$ for the branching ratios of LFV tau decays into hadrons in the SLH model. They are normalized to the present upper bounds in Table~\ref{tab:2}, i.e. a value of 1 in the y-axis indicates the present upper limit. We use $\chi_1 = 0.25$, $\chi_2 = 10 \, \chi_1$, $t_{\beta}=5$ and $\sin 2 \theta = 0.25$.} 
\end{figure}

 
In Figure~\ref{fig:4} we show the dependence on the scale of compositeness $f$ of the branching ratios (normalized to the upper bounds in Table~\ref{tab:2}) 
in the LFV hadron decays under study. We use $\chi_1 = 0.25$, $\chi_2 = 10 \, \chi_1$, $t_{\beta} = 5$ and $\sin 2 \theta = 0.25$. The plotted range for the
scale of compositeness seems the most natural in these models, however a higher value of $f$ might also make sense. In any case Figure~\ref{fig:4} shows
clearly the trend of the prediction. It can be seen that, in the
most optimistic case, for low values of $f$, our results imply branching ratios at least four orders of  magnitude smaller than present limits.
\par
%
\begin{figure*}[!]
\begin{minipage}{0.49\textwidth}
\centerline{\includegraphics[width=1.1\linewidth]{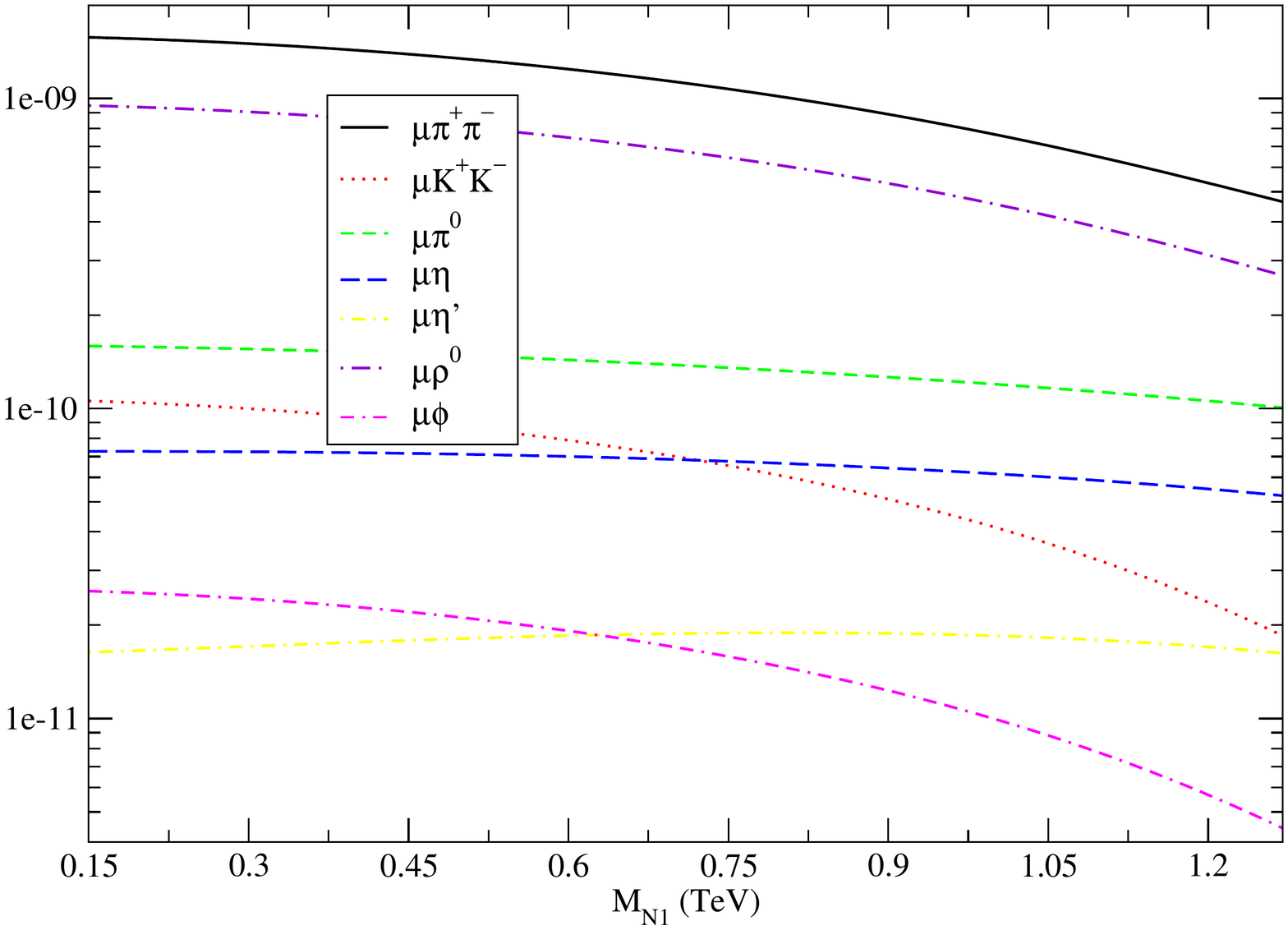}}
\end{minipage} 
\begin{minipage}{0.49\textwidth}
\centerline{\includegraphics[width=1.1\linewidth]{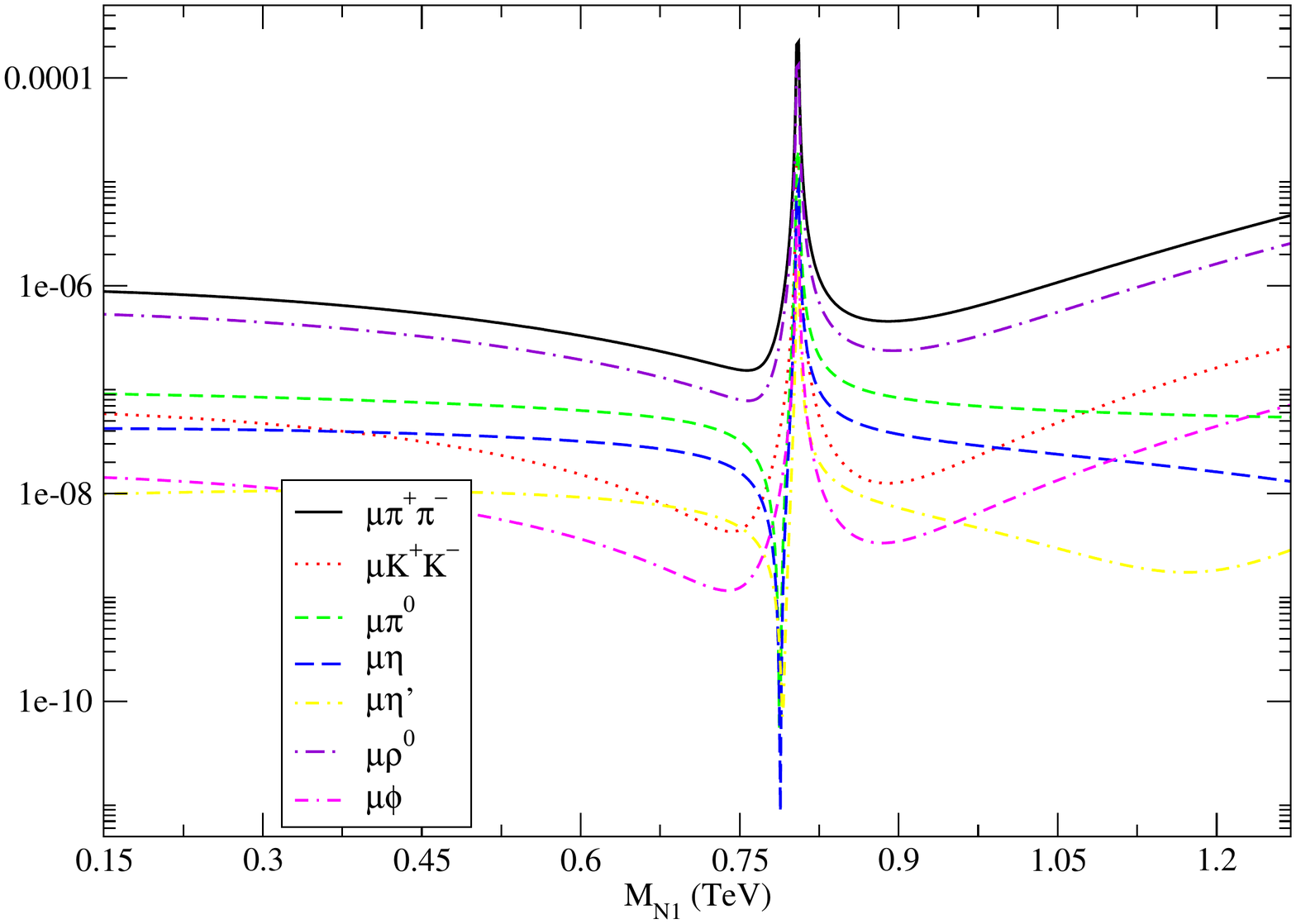}}
\end{minipage} 
\caption{\label{fig:5} Dependence of the LFV branching ratios on $M_{N_1}$. On the left we assume a small splitting of the heavy neutrino spectrum: $\chi_2 = 1.1 \, \chi_1$. On the right we assume a large splitting: $\chi_1 = 10 \, \chi_2$. We input $t_{\beta} = 5$, $\sin 2 \theta = 0.25$ and $f = 6 \, \mbox{TeV}$. Normalization as in Figure~\ref{fig:4}.}
\end{figure*}

%
The dependence on the \lq \lq little" neutrino masses is collected in Figure~\ref{fig:5}. We assume the cases of a small splitting: 
$\chi_2 = 1.1 \, \chi_1$ and a large one $\chi_2 = 10 \, \chi_1$. The structure that can be observed on the right hand side of Figure~\ref{fig:5}
is due to the effect produced by the large splitting in heavy neutrino masses when the second neutrino reaches and goes over the mass of the
heavy gauge boson $W'$. Naturally a small splitting produces branching ratios much smaller due to the unitarity of the lepton mixing matrix (see Eq.~(\ref{eq:sumneu2})).
%
%
\par 
In Figure~\ref{fig:5} we show the dependence of the branching ratios on the parameters $\tan  \beta$ and $\sin 2 \theta$. It can be seen that the
dependence on $t_{\beta}$ is rather mild for $t_{\beta} \gsim 3$. 
%
\begin{figure*}[!]
\begin{minipage}{0.49\textwidth}
\centerline{\includegraphics[width=1.1\linewidth]{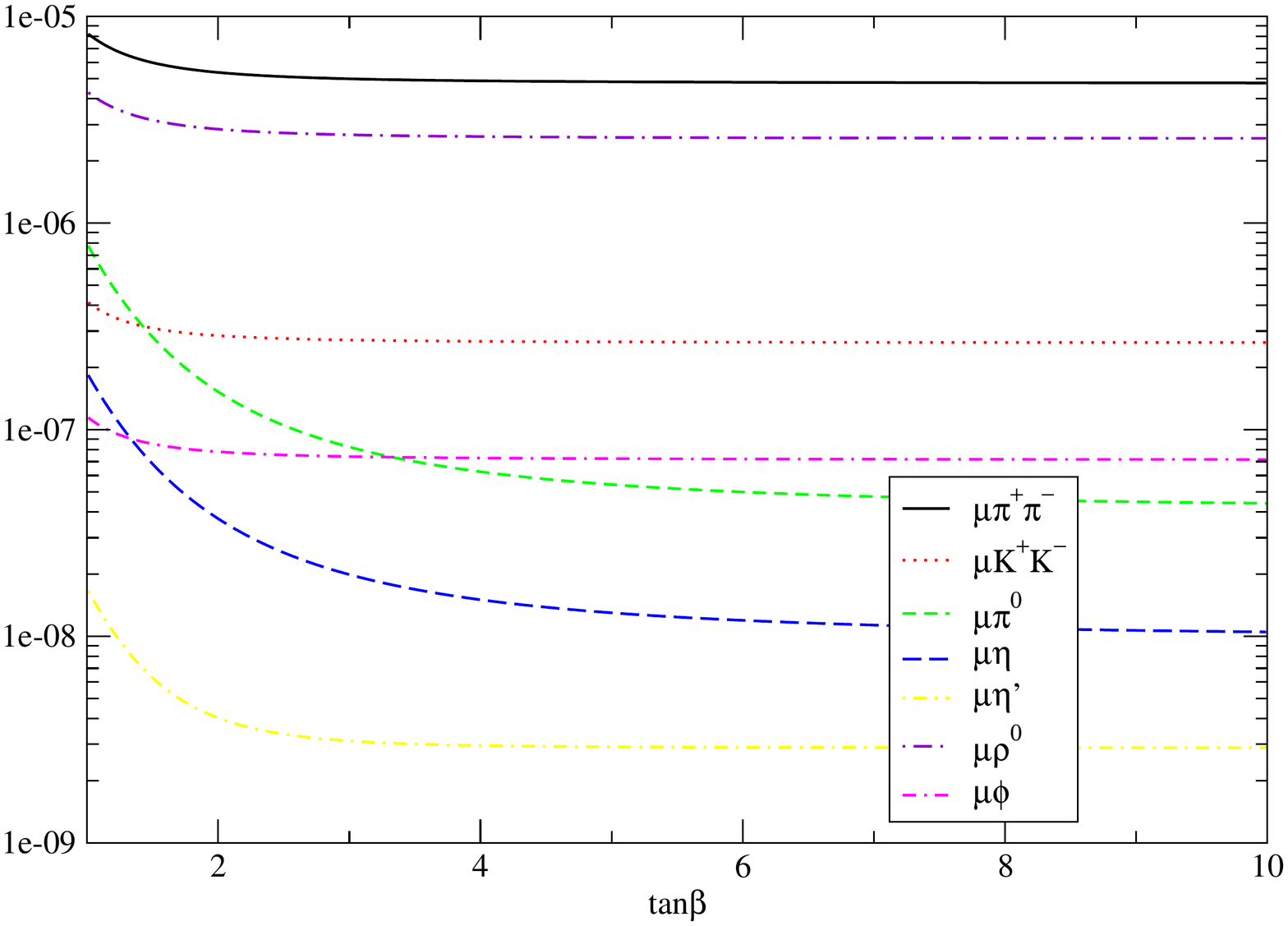}}
\end{minipage} 
\begin{minipage}{0.49\textwidth}
\centerline{\includegraphics[width=1.1\linewidth]{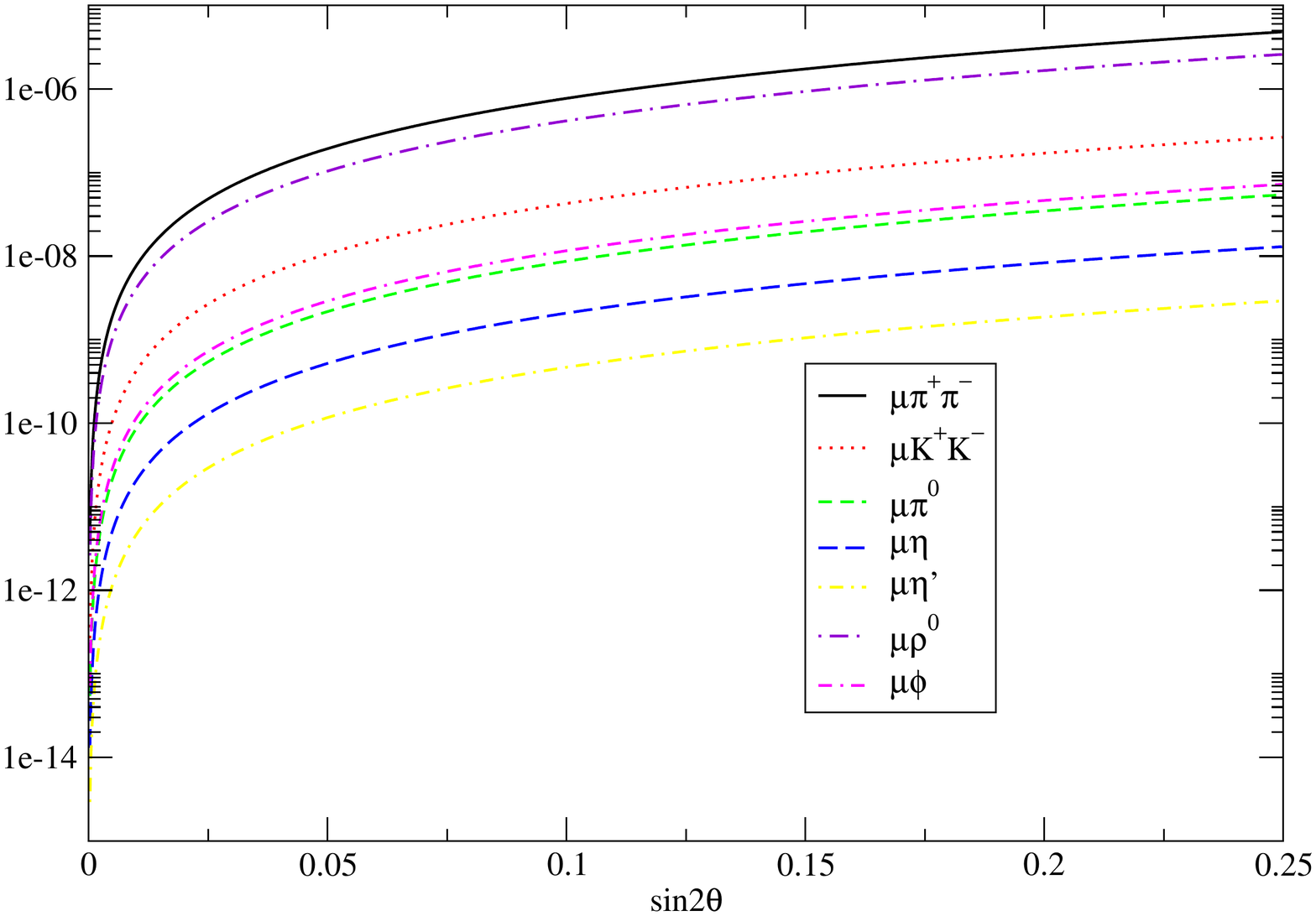}}
\end{minipage} 
\caption{\label{fig:6} Dependence of the LFV branching ratios on $\tan \beta$ (left) and $\sin 2 \theta$ (right). We input 
$(f,\sin 2 \theta) = (6 \, \mbox{TeV},0.25)$ on the left plot and $(f,t_{\beta}) = (6 \, \mbox{TeV}, 5)$ on the right plot. $\chi_1 = 0.25$, $\chi_2 = 10 \, \chi_1$ are used in both panels. 
Normalization as in Figure~\ref{fig:4}.}
\end{figure*}
%
\par
\begin{figure}[!]
\begin{minipage}{0.49\textwidth}
\centerline{\includegraphics[width=1\linewidth]{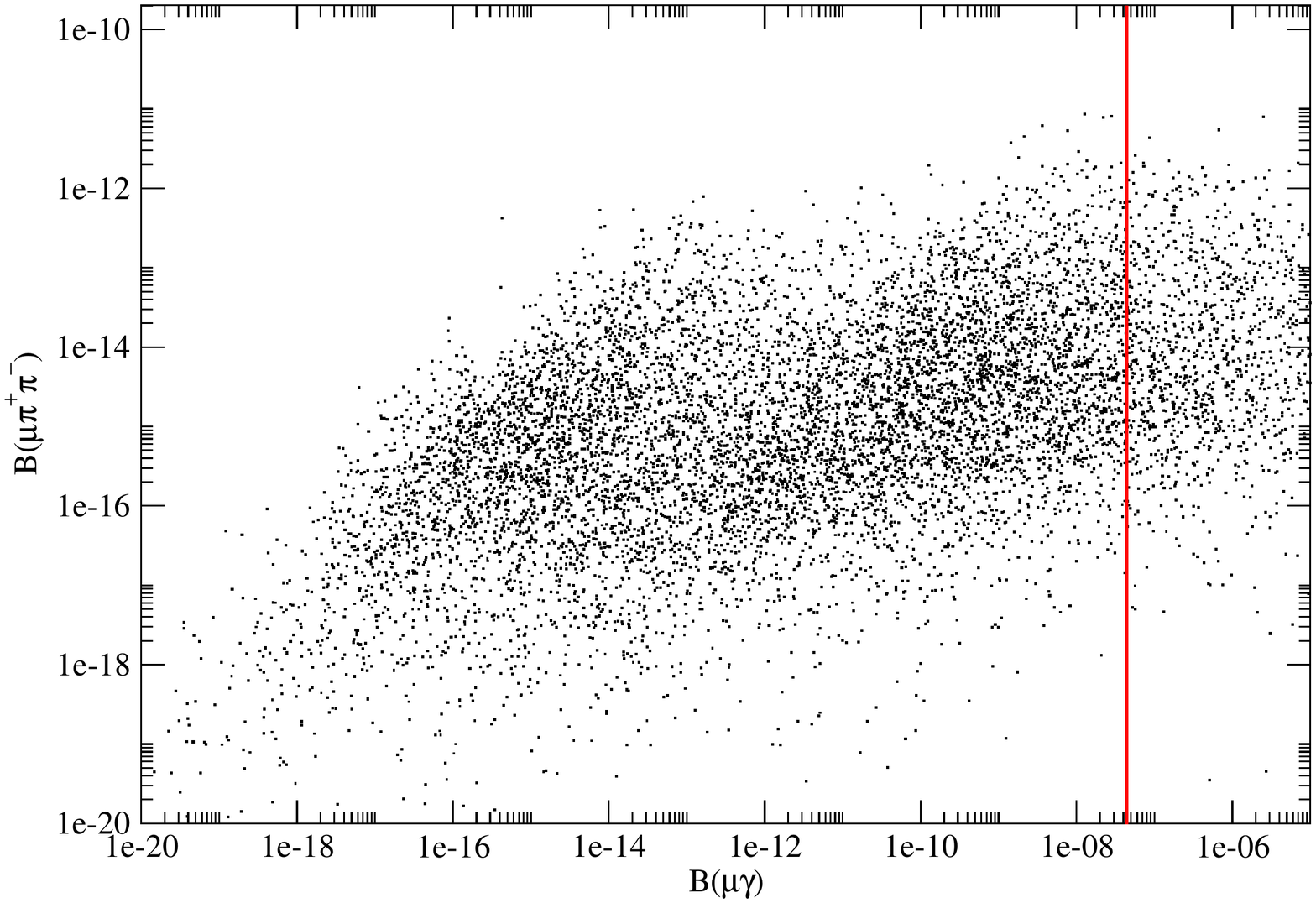}}
\end{minipage}  
\begin{minipage}{0.49\textwidth}
\centerline{\includegraphics[width=1\linewidth]{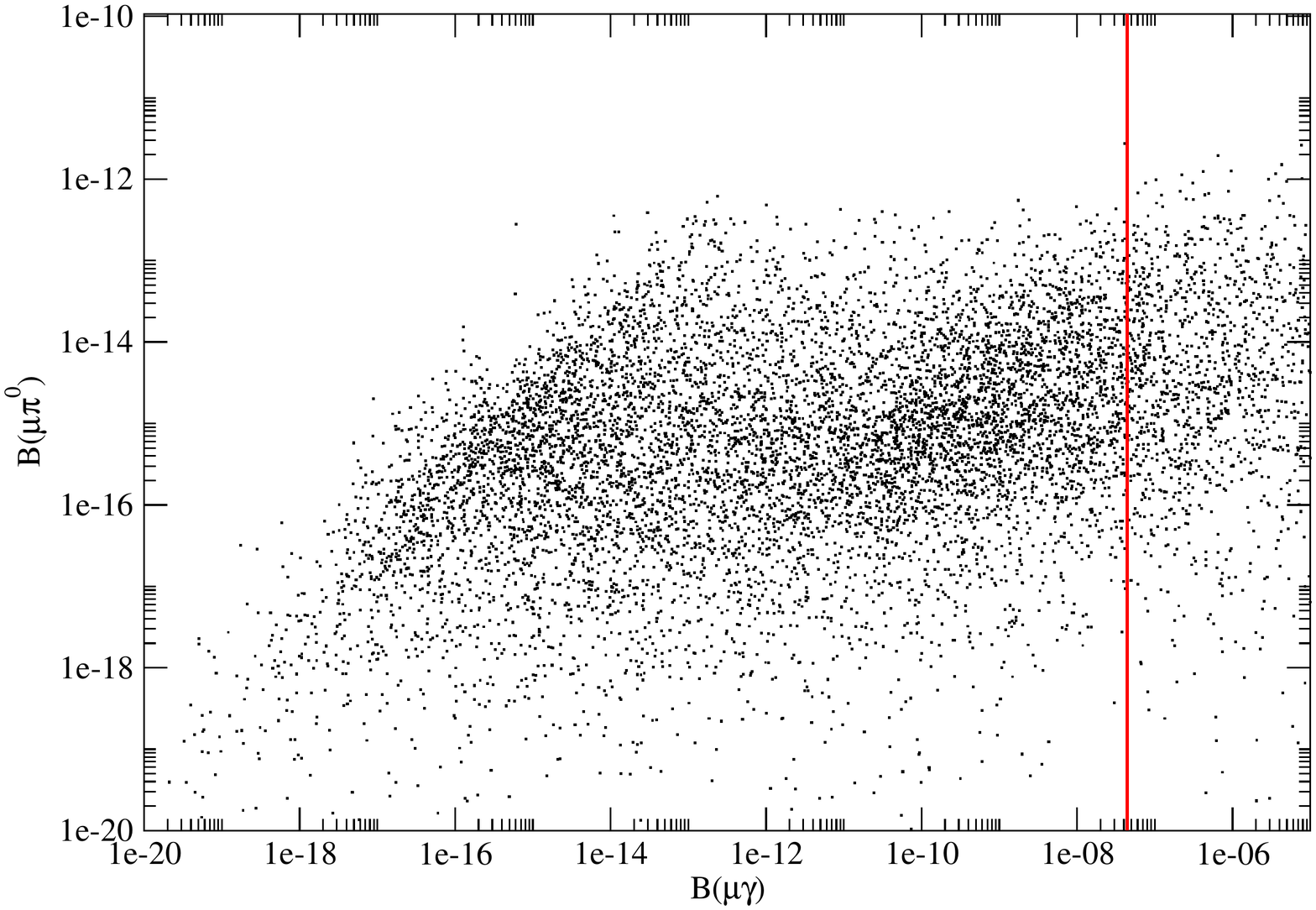}}
\end{minipage}  
\begin{minipage}{0.49\textwidth}
\centerline{\includegraphics[width=1\linewidth]{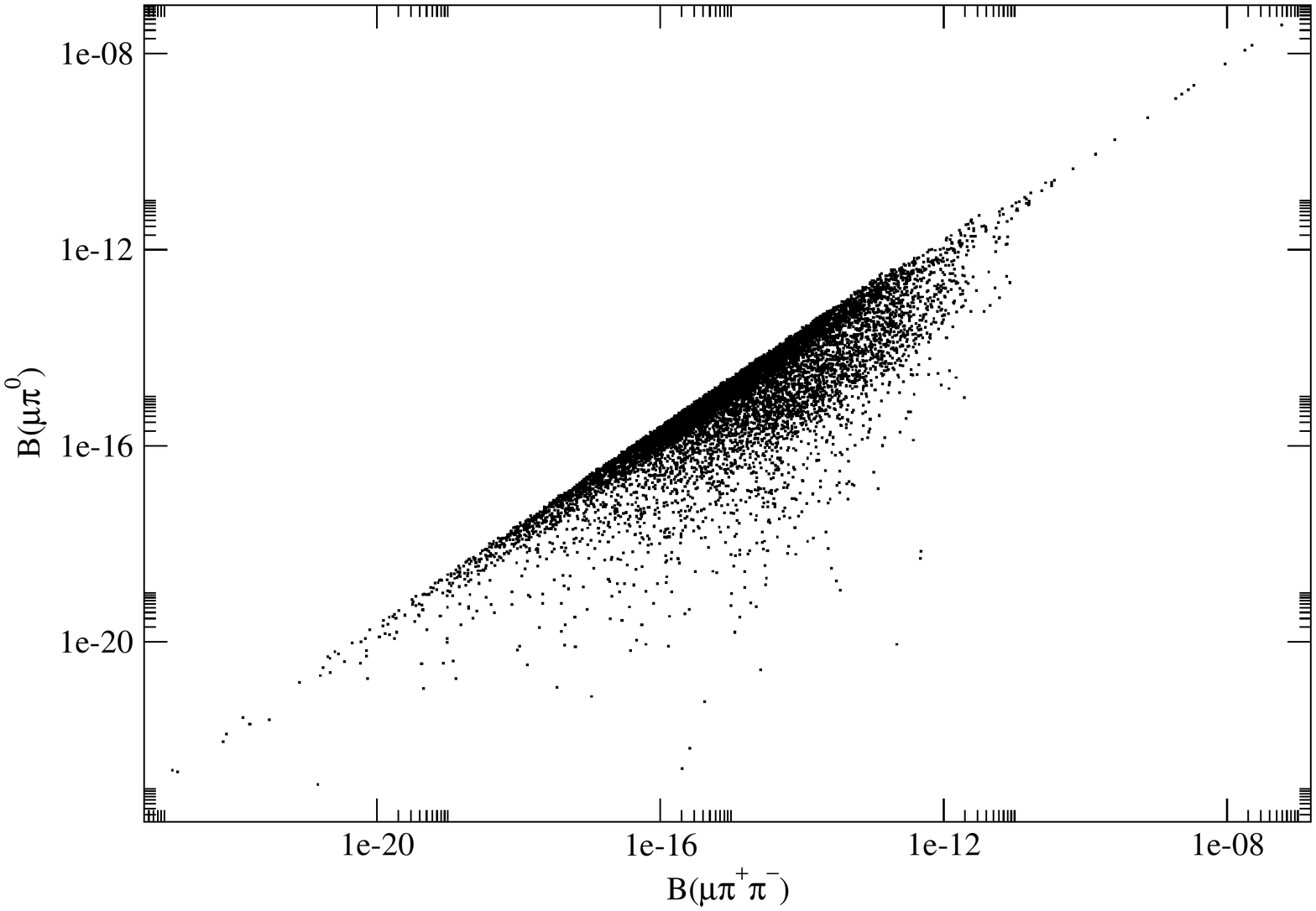}}
\caption{\label{fig:7} Scattered plots that show correlations between different branching ratios: $B(\tau \rightarrow \mu \pi^+ \pi^-)$ versus 
$B(\tau \rightarrow \mu \gamma)$ (upper plot), $B(\tau \rightarrow \mu \pi^0)$ versus $B(\tau \rightarrow \mu \gamma)$ (middle plot) and
$B(\tau \rightarrow \mu \pi^0)$ versus $B(\tau \rightarrow \mu \pi^+ \pi^-)$ (lower plot). We vary $f \in (2 \, \mbox{TeV}, 10 \, \mbox{TeV})$, 
$\sin 2 \theta \in (0,0.25)$, $t_{\beta} \in (1,10)$, $\chi_1\in (0,0.25)$ and $\chi_2 = a \chi_1$ for $a \in (1.1,10)$. Red lines indicate
the present upper bound for $B(\tau \rightarrow \mu \gamma)$ at $90 \% \, \mbox{C.L.}$.}
\end{minipage}  
\end{figure}

 
Correlations between different branching ratios are shown in Figure~\ref{fig:7}. The branching ratio of $\tau \rightarrow \mu \gamma$ has been obtained
from the SLH prediction for $\mu \rightarrow e \gamma$ in Ref.~\cite{delAguila:2011wk}. In this figure we have not normalized the branching ratios to the upper bounds as we did in previous figures. In the two first we show $B(\tau \rightarrow \mu \pi^+ \pi^-)$ and $B(\tau \rightarrow \mu \pi^0)$ versus $B(\tau \rightarrow \mu \gamma)$ and the
vertical red line indicates the measured present upper bound for the later decay. That would leave
hadron branching ratios, at the most, of ${\cal O}(10^{-12})-{\cal O}(10^{-14})$ as we already commented in the previous discussion. In the lower figure we plot $B(\tau \rightarrow \mu \pi^0)$ versus $B(\tau \rightarrow \mu \pi^+ \pi^-)$ and it is shown that both are highly correlated (we come back to this point later).
%
%
\par
\begin{figure}[h]
\centerline{\includegraphics[width=1.05\linewidth]{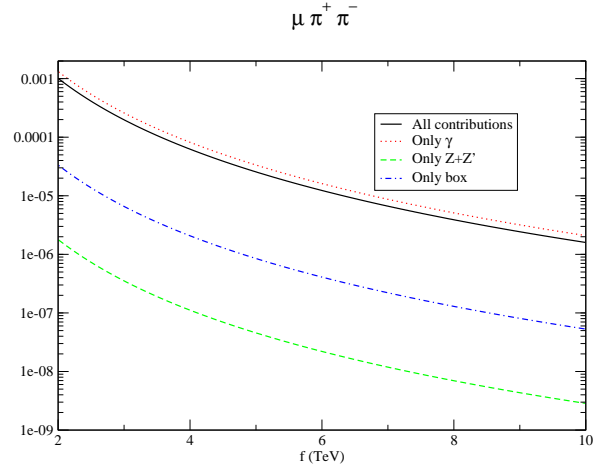}}
\caption{\label{fig:8} Dependence on the scale of compositeness $f$ for $\tau \rightarrow \mu \pi^+\pi^-$ showing the relative weights of the different contributions 
in the unitary gauge. Normalization and input as in Figure~\ref{fig:4}.} 
\end{figure}
\begin{figure}[h]
\centerline{\includegraphics[width=1.05\linewidth]{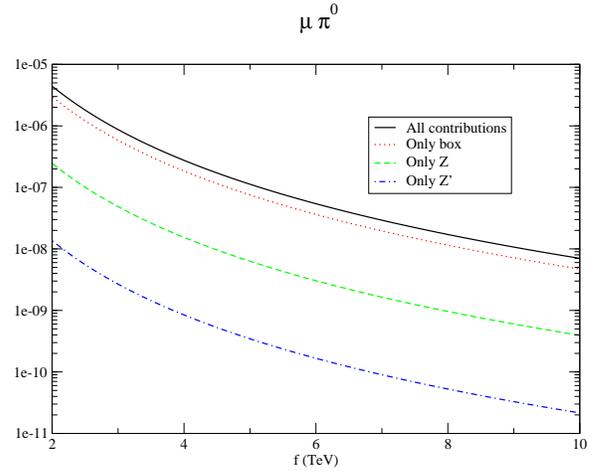}}
\caption{\label{fig:9} Dependence on the scale of compositeness $f$ for $\tau \rightarrow \mu \pi^0$ showing the relative weights of the different contributions 
in the unitary gauge. Normalization and input as in Figure~\ref{fig:4}.} 
\end{figure}

We would like to turn now to comment a property of our calculation in the SLH model. This is related with the relative weight of the different 
contributions that have been specified in Section~III. Supersymmetric scenarios seem to indicate that, in the 't Hooft-Feynman gauge, box diagrams
provide negligible contributions in comparison with photon-penguin diagrams in leptonic processes \cite{Hisano:1995cp,Arganda:2005ji}. However it has
been pointed out that this might not be the case in other models. For instance, in the Littlest Higgs model with T-parity  
both contributions are of the same order in  $\mu \rightarrow ee\overline{e}$  \cite{delAguila:2008zu} and the same happens in the same purely leptonic processes within the SLH model
\cite{delAguila:2011wk}. Notwithstanding in LFV hadron decays of the tau lepton within this later model, that we have studied in this article, we do not reach the same conclusion, at least in the unitary gauge. In Figure~\ref{fig:8} we show the dependence on $f$ for the different contributions for $\tau \rightarrow \mu \pi^+ \pi^-$. It can be seen that in hadron decays photon-penguin diagrams dominate over the rest of contributions. Box diagrams give a small input although one can see that they interfere destructively with the photon ones. Meanwhile the $Z$- and $Z'$-penguin diagrams are negligible.  
In Figure~\ref{fig:9} we plot the analogous comparison for $\tau \rightarrow \mu \pi^0$ where, obviously, there are not photon-penguin diagrams contributing. Then box diagrams give the bulk of the branching ratio though with a non-negligible positive interference of the Z-penguin contribution.
In the lower plot of Figure~\ref{fig:7} we noticed the high correlation between both $\tau \rightarrow \mu \pi^+ \pi^-$ and
$\tau \rightarrow \mu \pi^0$ decays. This seems eye-catching because, as we have seen, both processes are dominated by different contributions: the first by the photon-penguin diagrams and the second by the boxes. However, as it can be seen, the parameters of the model and the hadronization establish a correlation between
hadron processes that is not so apparent in $\tau \rightarrow \mu \gamma$, for instance.
%
%
%
%
\par 
As commented in Section~III we did not include Higgs-penguin contribution on the basis that their couplings to light quarks are suppressed by their masses. 
In Refs.~\cite{Celis:2013xja,Celis:2014asa} it was pointed out that a Higgs could couple, through a loop of heavy quarks, to two gluons able to hadronize 
into one or two pseudoscalars and, at least in the latter case, give a relevant contribution comparable with the one of the photon-penguin amplitude. 
This is indeed a two-loop calculation in our framework and we have not considered to sum this addition. In our opinion this could change our results for 
a factor not larger than ${\cal O}(1)$ and therefore it would not change our main conclusions. 
\par 
In Ref.~\cite{delAguila:2011wk} it was indicated that, at least in LFV decays of the muon into leptons and muon conversion 
in nuclei, the behaviour of the SLH model is very similar to the Littlest Higgs with T-parity. If that assertion could be extended to the hadron decays of the
tau lepton, as it seems rather sensible, we would definitely conclude that Little Higgs models predict a high suppression for these channels. It is now the turn of the flavour factories to clarify this issue.

\section{Conclusions}
We have analysed LFV decays of the tau lepton into one pseudoscalar, one vector or two pseudoscalar mesons in the Simplest Little Higgs model, characterized
as a composite Higgs model with a simple group $SU(3)_L \otimes U(1)_X$ and with a scale of compositeness $f \sim 1 \, \mbox{TeV}$ were a feature of 
collective symmetry breaking occurs providing a light Higgs boson. This model has interesting features like a reduced extension of the spectrum of gauge bosons and fermions over the SM ones and a small number of unknown parameters. In contrast the model has no custodial symmetry, though its lack does not 
bring large unwanted corrections. For the inclusion of the quark sector we use the {\em anomaly-free} embedding that does not need the role of an ultraviolet completion in order to cancel a gauge anomaly in the extended sector.  
The model has already been confronted with LHC data \cite{Reuter:2012sd,Reuter:2013zja} and keeps its strength waiting
for more precise determinations.
\par 
Lepton Flavour Violating decays are, due to their high suppression in the SM, an excellent benchmark where to look for new physics. Though present upper
bounds are very tight both in $\mu \rightarrow e \gamma$ and other muonic decays into leptons where one could expect that LFV, if any, will be first observed,
tau physics provide the unique property of being the only lepton decaying into hadrons and, consequently, offer a new scenario that, moreover, has been
thoroughly explored in B-factories like Babar and Belle. 
Present upper limits on branching ratios of the studied processes are of ${\cal O}(10^{-8})$ and future flavour factories, like Belle II, could lower those up to two orders of magnitude. Therefore the study of LFV hadron decays of the tau lepton is all-important in order to face the near future experimental status. 
\par 
We have considered the study of several hadron decays of the tau lepton, i.e. $\tau \rightarrow \mu  (P, V, PP)$ decays where $P$ is short for  a pseudoscalar meson and $V$ for a vector one. The leading amplitude for these decays, in the SLH model, is given by a one-loop contribution dynamically driven by the mixing
of the light charged leptons, $\tau$ and $\mu$ with the heavy \lq \lq little" neutrinos of the model. We have carried out the calculation at leading
order in the $v/f$ expansion and our results are ${\cal O}(v^2/f^2)$. For the numerical determination of the branching ratios we have considered previous constraints on the input constants of the model, although we have allowed their variation rather prodigally in order to convey the generic pattern of the predictions. Hence
we have studied the dependence of the branching ratios on the relevant parameters of the model.
\par 
We conclude that, for the most natural settings, the predictions of the SLH model for theses processes are, typically,
between 4 and 8 orders of magnitude smaller than present upper bounds and, therefore, out of reach for the foreseen next flavour factories. An observation of any of these decays by Belle II not only would signal new physics but also
would falsify the SLH model. 
\\

\begin{acknowledgments}
This research has been supported in part by the Spanish Government, Generalitat Valenciana and ERDF funds from the EU Commission [grants FPA2011-23778, FPA2014-53631-C2-1-P, PROMETEOII/2013/007, SEV-2014-0398]. A.L. also acknowledges Generalitat Valenciana for a Grisol\'{\i}a scholarship. J.P. would like to acknowledge the Mainz Institute for Theoretical Physics (MITP) for enabling him to complete this work. P.R. acknowledges financial support from project 236394 (Conacyt, Mexico) and the hospitality of IFIC, where part of this work was done.
\end{acknowledgments}

\providecommand{\href}[2]{#2}\begingroup\raggedright
\endgroup

\end{document}